\documentclass[a4paper,11pt]{article}

\usepackage[margin=2cm]{geometry}
\usepackage{times,latexsym,epsfig,epsf,graphicx,appendix,lineno, longtable,cite, psfrag, subfigure, fancyhdr, rotating}
\usepackage[centertags]{amsmath}

\usepackage[T1]{fontenc}
\usepackage[utf8]{inputenc}
\usepackage{authblk}

\usepackage[table, usenames, dvipsnames]{xcolor}
\usepackage{textcomp}
\usepackage{multirow}
\usepackage{lscape}
\usepackage{feynmp}
\usepackage{graphics}
\newcommand{\degree}{\ensuremath{^\circ}}
\newcommand{\neut}[1]{\ensuremath{\nu_{#1}}}
\newcommand{\numu}{\neut{\mu}}
\newcommand{\nue}{\neut{e}}

\newcommand{\goesto}{\ensuremath{\rightarrow}}
\newcommand{\dms}[1][NULL]{\ifthenelse{\equal{#1}{NULL}}{
	\ensuremath{\Delta m^2}}{\ensuremath{\Delta m^2_{#1}}}}
\newcommand{\sstt}[1][NULL]{\ifthenelse{\equal{#1}{NULL}}{
	\ensuremath{\sin^2 2\theta}}{\ensuremath{\sin^2 2\theta_{#1}}}}

\title{ T2K and beyond}

\author[1]{M.G. Catanesi \thanks{gabriella.catanesi@cern.ch}}

\affil[1]{INFN Sezione di Bari, Italy}

\begin{document}

\maketitle

\begin{flushright}
\vspace{-9cm}
T2K-TN-224-V5.1\\
\vspace{9cm}
\end{flushright}


\pagenumbering{arabic}

\begin{abstract}

This article presents the \emph{status of the art} of the T2K experiment and the measurements prospects
for the incoming years. After a brief description of the experiment, the most recent results will be illustrated.\
The observation of the  electron  neutrino appearance in a muon neutrino beam and the 
new high-precision measurements of the mixing angle $\theta_{13}$  by the reactor experiments have led to a 
re-evaluation  of the expected sensitivity  to the oscillation parameters, relative to what was given in the original T2K proposal. 
For this reason the new physics potential of T2K for $7.8 \times 10^{21}$ p.o.t. and for a data exposure 3 times larger than it's expected to be reachable with accelerator and beam line upgrades in 2026  before the start of operation of the next generation of long-baseline neutrino oscillation experiments will also be described  in the text.
In particular the last challenging scenario  opens the door to the possibility to obtain, under some conditions, a 3 $\sigma$ measurement excluding  $\sin(\delta_{CP})= 0$.\\ 

\end{abstract}

\newpage
\tableofcontents
\newpage
\section{Introduction}
\label{intro}
In the last 15 years several experimental measurements confirmed the neutrino oscillations hypothesis by attacking the problem on different fronts:\\
\begin{itemize}

\item The observation of a zenith-angle-dependent deficit in muon neutrinos produced by high-energy proton interactions in the atmosphere \cite{1} supported the possibility that a particular
neutrino flavor can be transmuted into another one.
\item The \emph{anomalous} solar neutrino flux \cite{2} problem was shown to be due to neutrino oscillation by precise measurements \cite{3,4,5,6}. 
\item Taking advantage of nuclear reactors as intense sources, the disappearance of electron anti-neutrinos has been firmly established \cite{6,8,9,10}. 
\item The development of high-intensity proton accelerators that can produce focused neutrino beams with mean energy from a few hundred MeV to tens of GeV has enabled measurements 
of the disappearance of muon-neutrinos (and muon anti-neutrinos) \cite{7, 11, 12} 
and appearance of electron-neutrinos (and electron antineutrinos) \cite{13,14,15, 16} 
and tau-neutrinos \cite{17} over distances of hundreds of kilometres.

\end{itemize}

As a matter of fact the possibility for a  neutrino of a particular flavor to be transmuted into another flavor has profound implications demonstrating that neutrinos have mass. 
Recently this extraordinary achievement in physics was fully recognized in the scientific community by the awarding of the Nobel Prize 2015 to T. Kajita and A. B. McDonald. \\ 
To date, all the experimental  results cited above are well-described within the three neutrino generations  oscillation framework. 
In this case the unitary mixing matrix, often referred to as the PMNS (Pontecorvo-Maki-Nakagawa-Sakata) matrix \cite{18}, can be written as  a $3\times3$ matrix:

\begin{equation}
\begin{split}
U=\left(
\begin{array}{ccc}
 1 & 0  & 0  \\
 0 & +c_{23}  & +s_{23}  \\
 0 &  -s_{23} &   +c_{23}
\end{array}
\right)
\left(
\begin{array}{ccc}
 +c_{13} & 0  & +s_{13}e^{-i\delta_{CP}}  \\
 0 & 1  & 0  \\
 -s_{13}e^{i\delta_{CP}} & 0  & +c_{13}  
\end{array}
\right)
\left(
\begin{array}{ccc}
 +c_{12} & +s_{12}  & 0  \\
 -s_{12} & +c_{12}  & 0  \\
 0 & 0  &   1
\end{array}
\right)\\
\\
= \left(
\begin{array}{ccc}
c_{12}c_{13}  & s_{12}c_{13}  & s_{13}e^{-i\delta_{CP}}  \\
 -s_{12}c_{23}-c_{12}s_{23}s_{13}e^{i\delta_{CP}} &c_{12}c_{23}-s_{12}s_{23}s_{13}e^{i\delta_{CP}}   & s_{23}c_{13}  \\
  s_{12}s_{23}-c_{12}c_{23}s_{13}e^{i\delta_{CP}}& -c_{12}s_{23}-s_{12}c_{23}s_{13}e^{i\delta_{CP}}  & c_{23}c_{13}   
\end{array}
\right)
\label{eq:mnsp}
\end{split}
\end{equation}

where $c_{ij}=\cos\theta_{ij}$ and $s_{ij}=\sin\theta_{ij}$. In this case four parameters are required to describe the matrix: three angles $\theta_{12}$, $\theta_{23}$ and $\theta_{13}$ and the CP violating phase $\delta_{CP}$.\\

In this context  the T2K experiment, proposed in 2003~\cite{19} and 
approved in 2006 to collect data corresponding to $7.8\times 10^{21}$ protons-on-target (p.o.t) from a 30 GeV proton beam at the JPARC accelerator facility in Japan, 
played until an important role, and will keep playing it in the future.\\

T2K is a long-baseline neutrino oscillation experiment designed to achieve the following main physics goals:

\begin{itemize}
\item 
search for \numu{}\goesto{}\nue{} appearance and establish  
 $\theta_{13} \ne 0$ with a sensitivity 
down to $\sstt{}_{13} \sim 0.008  (90\%~\mathrm{C.L.})$;
\item precision measurement of oscillation parameters in \numu{} disappearance mode
with $\delta{}(\dms{}_{32}) \sim 10^{-4}$~eV$^2$ 

and $\delta{}(\sstt{}_{23}) \sim 0.01$ ; 

\item search for exotic physics including Lorenz violation, search for sterile components in \numu{}\ ($\nu_e$) disappearance  at the near detector and more.
\end{itemize}

The T2K experiment began the physics data taking in 2010~\cite{20}.\
Despite the devastating March 2011 earthquake in eastern Japan, which caused severe damage to the accelerator complex at J-PARC, and abruptly discontinued the data-taking run of the experiment, in July 2011 the T2K collaboration announced a first indication of $\theta_{13} \ne 0$\cite{13} and in 2013  reached a major physics goal: the discovery of \numu{}\goesto{}\nue{} appearance at 7.3 $\sigma$ level of significance when a mere 8.4\% 
of the total approved p.o.t. \cite{15} was recorded. 
This is the first time that explicit flavor appearance has been observed 
from another neutrino flavor with a significance larger than 5$\sigma$.  This
observation opened the door to the search for CP violation (CPV) in the lepton sector.

Following this discovery, the primary physics goal for T2K and - at large -
for the neutrino physic community has become a detailed investigation 
of the three-flavor paradigm; this requires the
determination of the unknown CP-violating phase $\delta_{CP}$, 
the resolution of the mass hierarchy (MH), and the
precise measurement of $\theta_{23}$ to determine the octant.

T2K, along with the Nova \cite{21} experiment 
that started data taking one year ago, will play a major role in the determination of these parameters for at least a decade.\\

In this article, after a brief description of the experiment in Sec.\ref{sec:T2K},  
we will describe the most recent results obtained by T2K (Sec.\ref{sec:T2K_res}). 
In Sec.\ref{sec:T2K_78} we will provide a re-evaluation  of the expected sensitivity 
to the oscillation parameters, relative to what was given in the original T2K
proposal \cite{19}, by taking into account the actual observation of the  electron 
neutrino appearance in a muon neutrino beam by T2K and the new high-precision measurements of the mixing angle $\theta_{13}$ from reactor experiments. In  Sec.\ref{sec:T2K_250}.1 we will briefly describe 
the proposed upgrade plan for the J-PARC accelerators and neutrino experimental facility aiming to reach a 1.3 MW beam power \cite{37,38}. Finally in Sec.\ref{sec:T2K_250}.2 we will describe the physics potential of T2K for a data exposure 3 times larger than  the presently approved one. \\


\section{The T2K experiment}
\label{sec:T2K}
\begin{figure}[h!]
\begin{center}
\psfig{figure=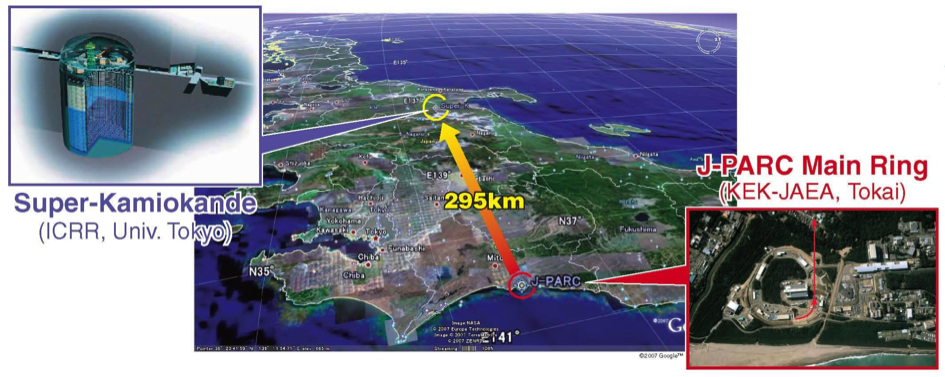,width=16cm}
\caption{The neutrino beam journey in the T2K experiment. The high intensity neutrino beam is produced at J-PARC (Tokai, Mito, Ibaraki Prefecture, Japan) and travels 295 km to the Super-Kamiokande detector (mount Kamioka, close to Hida, Gifu Prefecture, Japan).}
\label{fig:t2k_baseline}
\end{center}
\end{figure}

The T2K (Tokai-to-Kamioka) experiment is a second generation LBL (Long Base-Line) neutrino oscillation experiment to probe physics beyond the Standard Model. T2K uses an almost pure $\nu_\mu$ beam produced using the new MW-class proton synchrotron at J-PARC\footnote{Japan Proton Accelerator Research Complex jointly constructed and operated by KEK and JAEA}. The neutrino beam is detected first in the near detector \emph{ND280} and then travels 295 km to the far detector \emph{Super-Kamiokande} (see Fig.\ref{fig:t2k_baseline}).\\

\begin{figure}[h!]
\begin{center}
\psfig{figure=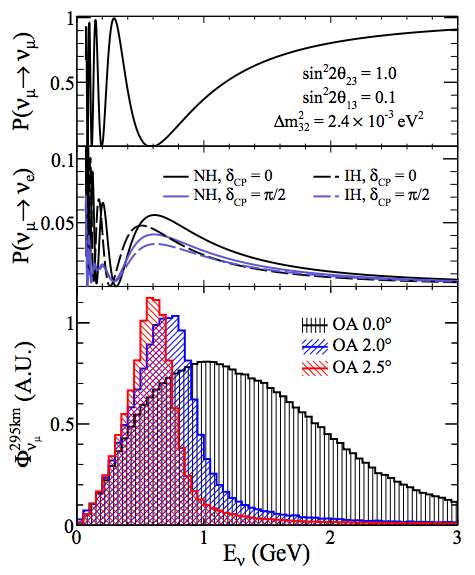,width=8cm}
\caption{The muon neutrino survival probability (top) and electron neutrino appearance probabilities (middle) at 295 km, and the un-oscillated neutrino fluxes for different values of the off-axis angle (OA) (bottom). The appearance probability is shown for two values of the phase $\delta_{CP}$, and for normal (NH) and inverted (IH) mass hierarchies.}
\label{fig:t2k_off}
\end{center}
\end{figure}

T2K adopts the off-axis method to generate a narrow-band neutrino beam using the proton synchrotron at J-PARC. In this method the neutrino beam is purposely directed at an angle with respect to the baseline connecting the proton target to the far detector, Super-Kamiokande. The off-axis angle is set at $2.5\degree$ so that the narrow-band $\nu_\mu$ beam generated towards the far detector has a peak energy at $\sim$0.6 GeV (see figure \ref{fig:t2k_off}). Such configuration maximizes the effect of the neutrino oscillation at 295 km and minimizes the background to electron-neutrino appearance detection.  The angle can be reduced to $2.0\degree$, allowing variation of the peak neutrino energy, if necessary. The J-PARC beamline can also provide to the experiment an anti-neutrino beam instead of a neutrino beam. As it will be shown in  sections \ref{sec:T2K_78} and \ref{sec:T2K_250}, this aspect is very important to constrain the $\delta_{CP}$ phase. 

\subsection{Near Detectors}

The near detectors were constructed in an underground  hall of 33.5m depth and
17.5m diameter at 280m downstream from the target.
They are used to measure the neutrino energy spectrum, flavor content, and interaction rates of the un-oscillated beam, and to predict the neutrino interactions at Super-Kamiokande.
Two detectors were installed; an on-axis detector (aimed in the direction of the neutrino beam center), and an off-axis detector (aimed in the direction of SK).\\

\begin{figure}[h!]
\begin{center}
\psfig{figure=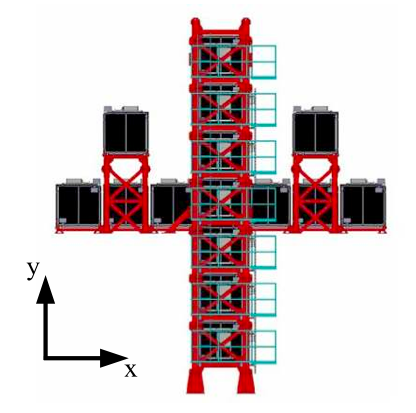,width=8cm}
\psfig{figure=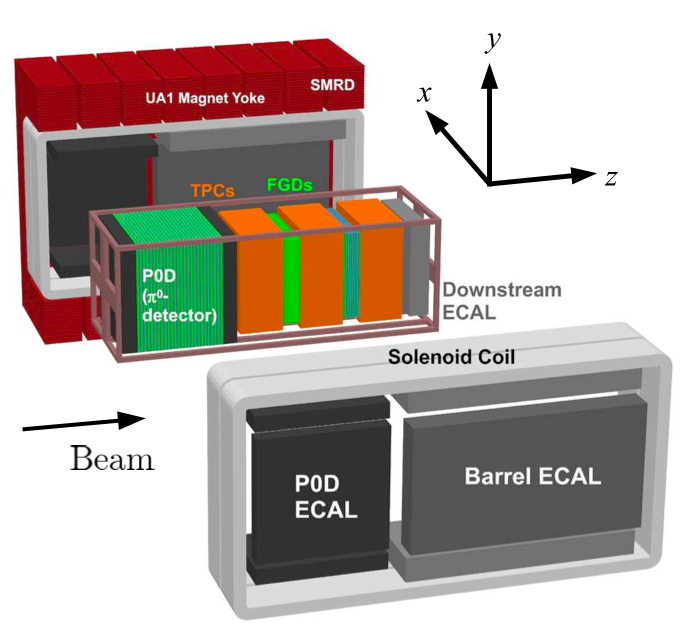,width=8cm}
\caption{Schematic view of the INGRID on-axis detector (left), and the ND280 off-axis detector (right).}
\label{fig:near_det}
\end{center}
\end{figure}

The on-axis detector, namely the INGRID detector, consists of 16 1mx1mx1m cubic modules
as shown in Fig.\ref{fig:near_det} (left). Each module is a "sandwich" of iron/scintillator detectors: 14 of them are arranged so as to form a cross configuration, 
and the remaining two diagonal modules are positioned off the cross axes.\
The center of the cross corresponds to the neutrino beam center, defined as $0\degree$ with respect to the direction of the primary proton beamline. INGRID is able to measure the on-axis neutrino beam profile and direction with an accuracy of $\pm$ 1 mrad. 
Note that the monitoring of beam direction with high precision is very important for the \emph{off-axis} configuration. In fact a 1 mrad divergence corresponds to a change of 2$\%$ in the integral of the neutrino flux expected at SK. \\  

The off-axis detector (named ND280) shown in Fig.\ref{fig:near_det} (right) is a magnetized tracking detector. The detector elements are contained inside the refurbished UA1 magnet. Inside the upstream end of magnet sits a $\pi^0$ detector (P$\emptyset$D) consisting of tracking planes of scintillating bars alternating with either water target/brass foil or lead foil. Downstream of the P$\emptyset$D is the tracker, composed by three time projection chambers (TPCs) and two fine grained detectors (FGDs) consisting of layers of finely segmented scintillating bars.\\
 The tracker is designed to study  charged current interactions in the FGDs. The P$\emptyset$D, TPCs, and FGDs are all surrounded by an electromagnetic calorimeter (ECal) for detecting electrons and photons to better constrain the $\nu_e$ contamination in the beam and the $\gamma$ background, while the return yoke of the magnet is instrumented with scintillator slabs to measure the range of the muons (Side Muon Range Detectors, SMRD) escaping from the sides of the off-axis detector.\\
In addition, the near off-axis detector can also perform accurate cross-section measurements on different target materials (carbon, water, oxygen, argon) to minimise the impact of the interaction model uncertainties on the systematic error budget. 
 
\subsection{Super-Kamiokande}

The far detector, Super-Kamiokande (SK), is a 50 kton water Cherenkov detector
 \cite{22} located 1000 m underground in the Kamioka mine, Japan. Its distance from J-PARC is 295km. \
In the inner detector (ID), 22.5kton of fiducial volume are viewed by 11,129 20-inch diameter PMTs. 
The outer detector (OD), which surrounds the ID, is also a water Cherenkov detector:
it is used to veto events that enter or exit the ID. SK started its operation in April 1996. After a complete upgrade of its electronics systems in 2008, it was named SK-IV.
The most important characteristic of SK, as the far detector of the T2K experiment, 
is its ability to differentiate between muons and electrons with high efficiency.  
Considering that the majority of the neutrino interactions in this range of energies are charged current quasi-elastic (CCQE) interactions, 
the identification of muons and electrons directly implies the identification of the parent $\nu_\mu$ ($\bar{\nu}_\mu)$ or $\nu_e$ ($\bar{\nu}_e)$.
Details about the SK particle identification performances are reported in \cite{23}. It was verified that the probability of  misidentification is less than 1 $\%$.


%

\subsection{Data Taking Status}
\begin{figure}[h!]
\begin{center}
\psfig{figure=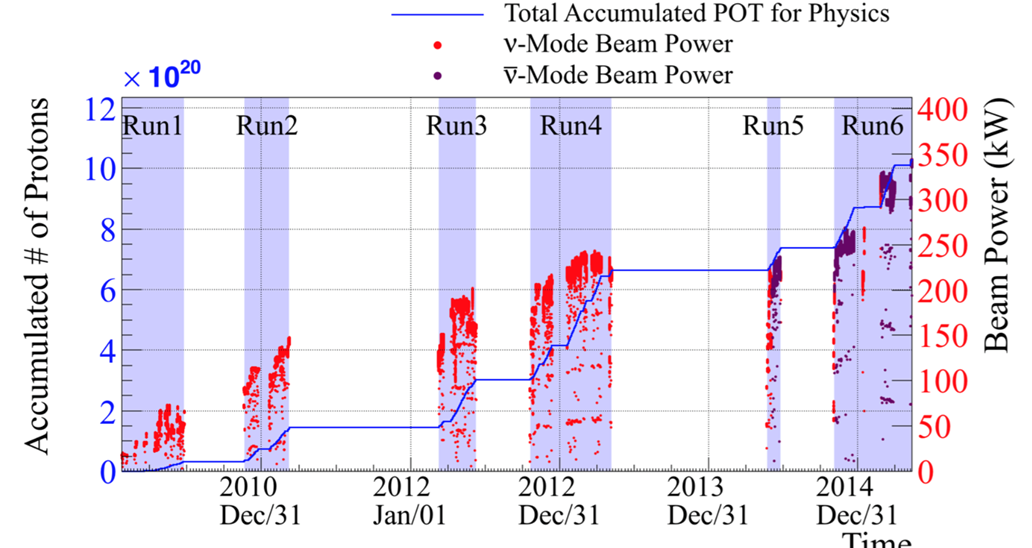,width=16cm}
\caption{The evolution of the primary proton beam intensity and integrated proton delivery  in the T2K experiment. Red (violet) dots show averaged beam power per hour in the neutrino (anti-neutrino) mode beam; the scale is given in the right vertical axis. The blue solid line shows the accumulated number of delivered protons from the beginning of the experiment. The scale is given in the left vertical axis.}
\label{fig:DataTaking}
\end{center}
\end{figure}

The evolution of the proton beam delivery is shown in Fig.\ref{fig:DataTaking}. The physics data-taking started in January 2010. That year one beam pulse had only six bunches 
in 5 microseconds. The number of protons per pulse (ppp) was limited to  $2.\times 10^{13}$, and the beam pulse cycle was at that time 3.52 seconds.
Many efforts were made by the J-PARC accelerator group to increase the beam power.
To date the number of bunches in each pulse is 8, and the number of protons per pulse 
$1.8\times 10^{14}$. In the mean time, the beam cycle time was reduced to 2.48 seconds. 
The maximum beam power achieved through June 2015 was 371 kW.
T2K accumulated $1.1\times 10^{21}$ p.o.t. data until June 4, 2015. 
This is about 14$\%$ of  the final goal of $7.8\times 10^{21}$ p.o.t., which can be obtained 
over five years of beam operation \cite{36,37,38}.\\
In June 2014, the direction of the magnetic horn current was reversed and an anti-neutrino
beam run was started.\
In the next section, results based on $6.57\times 10^{20} $p.o.t. neutrino beam data
\cite{12, 15, 24}  and $4.04\times 10^{20} $ p.o.t. anti-neutrino beam data  recorded until June 2015 will be reported \cite{25}.

\newpage

\subsection{Analysis Strategy }

T2K employs various analysis methods to estimate oscillation parameters.
In general, these methods extract oscillation parameters
from the data by comparing the observed and
predicted $\nu_e$ and $\nu_\mu$ interaction rates and 
energy spectra at the far detector. 
The rate and spectrum depend on the oscillation parameters, the incident neutrino flux,
neutrino interaction cross sections, and the detector response.
The initial estimate of the neutrino flux is determined by detailed
simulations incorporating proton beam measurements, INGRID measurements,
and the pion and kaon production rates measured by the 
NA61/SHINE~\cite{26,27} experiment.\\ 
The ND280 detector measurement of  $\nu_\mu$
charged current (CC) events (un-oscillated spectra) is used to constrain the initial flux estimates and
parameters of the neutrino interaction models that affect the
predicted rate and spectrum of neutrino interactions at both ND280 and Super-K.\\

\begin{figure}[h!]
\begin{center}
\psfig{figure=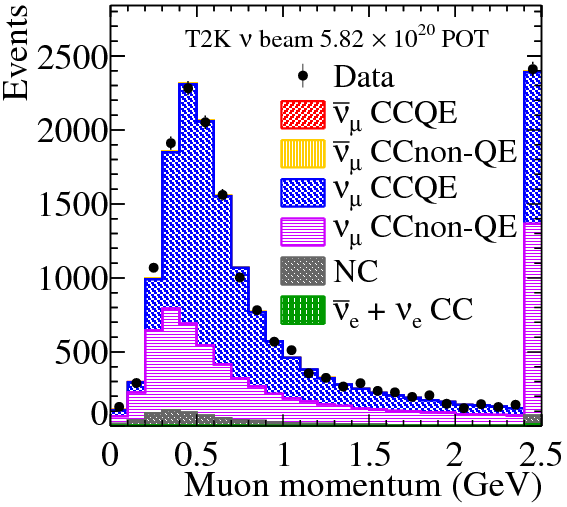,width=8cm}
\psfig{figure=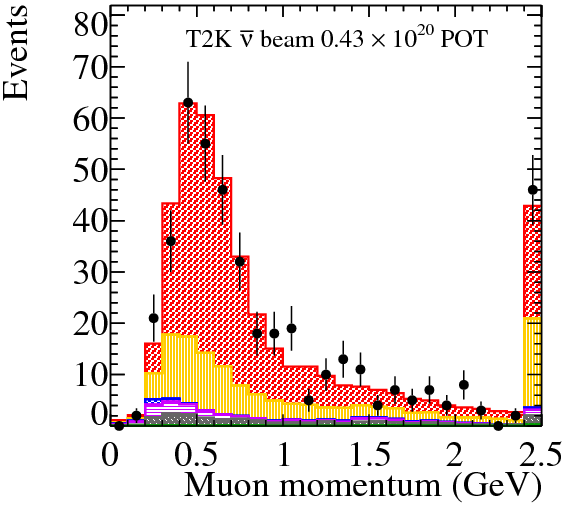,width=8cm}
\caption{  Muon Momentum distribution of CC   $\nu_\mu$  events  (left) and CC  $\bar{\nu}_\mu$ events (right) in ND280.
 Data points include statistical and systematic errors. After the adjustment of systematic uncertainties, there is agreement between the data and Monte Carlo simulation.}
\label{fig:ND280cc0pi}
\end{center}
\end{figure}

\

\begin{figure}[h!]
\begin{center}
\psfig{figure=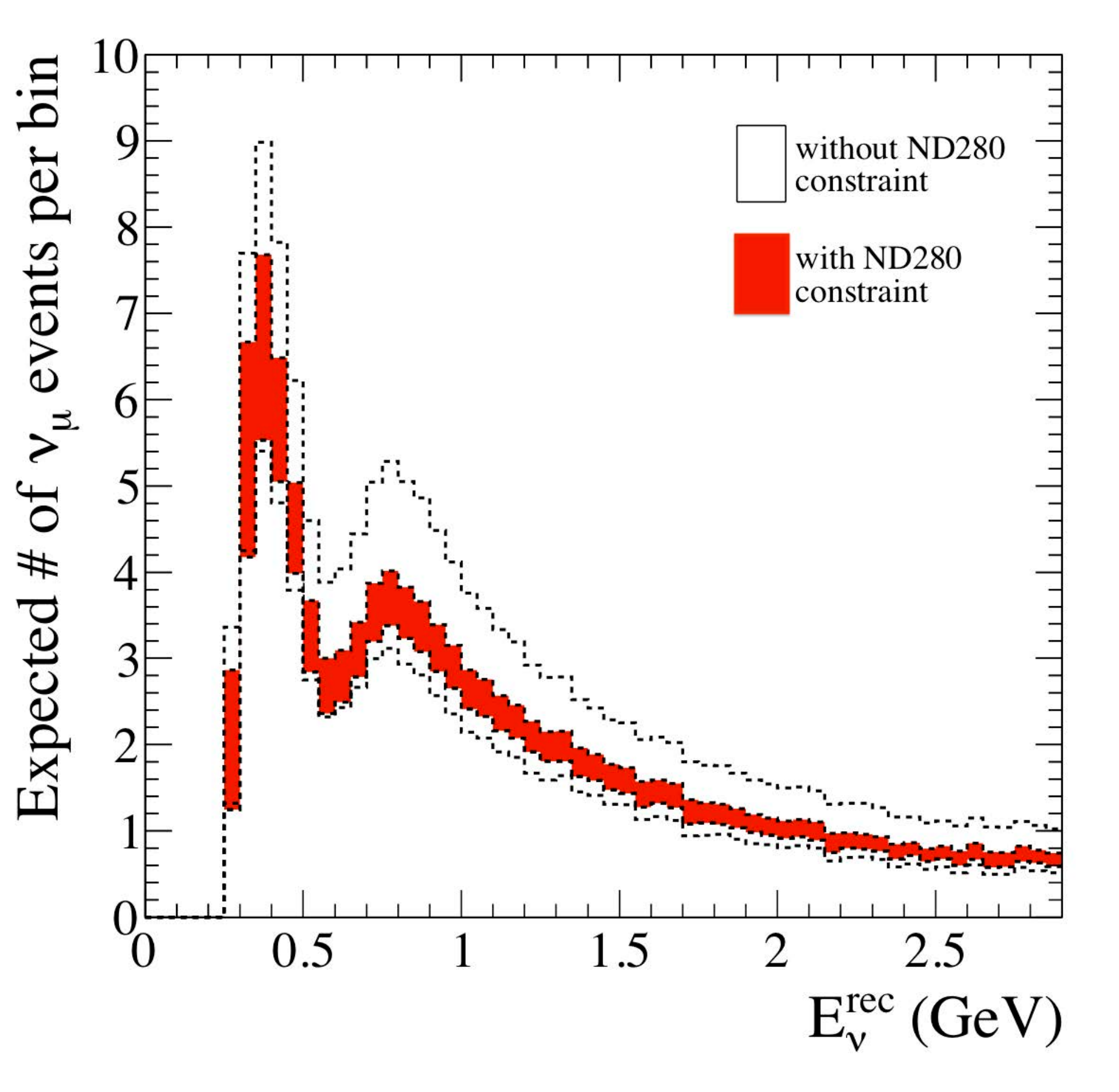,width=8cm}
\psfig{figure=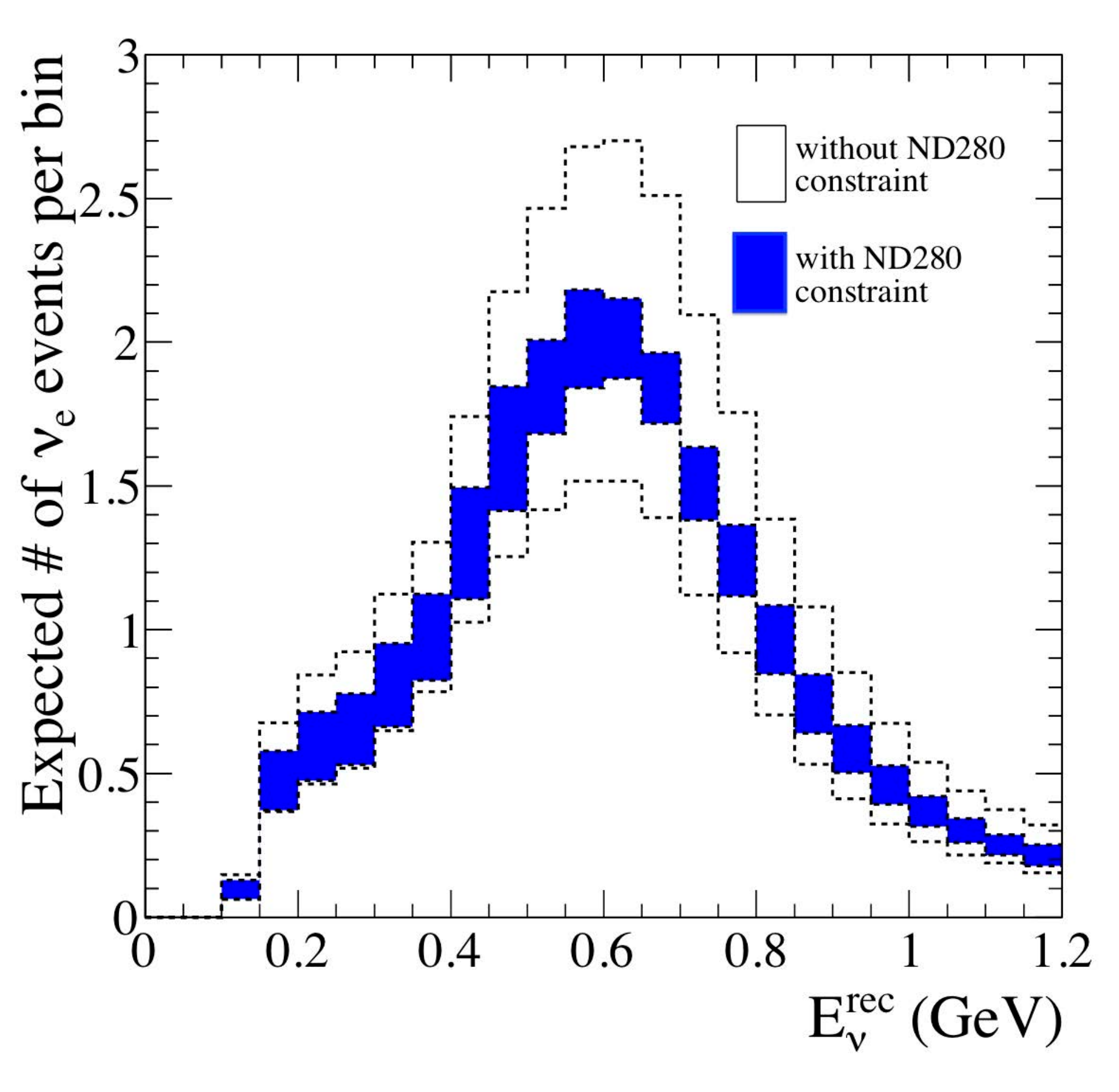,width=8cm}
\caption{ Total error envelopes for the reconstructed energy distributions of $\nu_\mu$ CC (left)
and $\nu_e$ CC (right) candidate events at SK , using typical oscillation parameter values, with and
without the ND280 constraint applied.}
\label{fig:SKSysErr}
\end{center}
\end{figure}

$\nu_\mu$ ($\bar{\nu}_\mu$) CC interactions are selected by requiring that the highest-momentum negative(positive)-curvature track in an event starts within the
upstream FGD (FGD1) fiducial volume (FV) and has an energy deposit in the middle TPC (TPC2)
 consistent with a muon. The muon  PID requirement is  based on a truncated mean
of  measurements  of  energy  loss  in  the  TPC  gas \cite{41}.
 Events with a track in the TPC upstream of FGD1 are rejected and the remaining 
 $\nu_\mu$ CC candidates are divided into three sub-samples according to the number of associated pions:
  $\nu_\mu$ CC 0$\pi$, dominated by CCQE interactions,
 $\nu_\mu$ CC 1$\pi^+$, dominated by CC resonant pion production, 
 and the so-called $\nu_\mu$ CC \textit{other}, dominated by deep inelastic scattering.\\
 Several control samples are used to assess the uncertainty in the modeling
of FGD and TPC response. A detailed description of the systematic errors considered in the analysis and the numerical evaluation of each of them can be found in \cite{13}.
  All the parameters related to cross sections and neutrino fluxes are adjusted based on the comparison between ND280 data and Montecarlo. As it is shown in Fig. \ref{fig:ND280cc0pi}, after the 
adjustment \cite{12,13}, the agreement is good. \\
Thanks to the inputs from the ND280 analysis, the systematic errors on the expected neutrino events at SK are strongly reduced. The effect is illustrated in Fig. \ref{fig:SKSysErr}: they decrease  from
23.5$\%$ to 7.7$\%$  for  $\nu_\mu$ candidates, and from 26.8\% to 6.8\% for $\nu_e$ candidates.\\

At SK,  $\nu_e$ and/or $\nu_\mu$ charged current quasi-elastic (CCQE) events are selected, with efficiencies and backgrounds determined through detailed 
simulations tuned to control samples, accounting for final state interactions (FSI) inside the nucleus and secondary hadronic interactions (SI) in the detector material.
These combined results are used in a fit to determine the oscillation parameters.\\

\newpage

\section{Recent T2K results}
\label{sec:T2K_res}
\subsection{Neutrino events selection in SK}
In the years 2010-2013, neutrino events corresponding to $6.57\times 10^{20}$ p.o.t. have been recorded in SK.
The event selection process comprises two steps. The first is the same for 
$\nu_\mu$ and $\nu_e$, and it allows to accept only the beam-related Fully-Contained Fiducial Volume (FCFV) events. 
Note that, with the addition of the request that the event time stamp is within a range of 2 to 10 $\mu$s from the beam spill time recorded in Tokai, 
the applied cuts are the same as in the well-established
atmospheric neutrino analysis \cite{28}.\\
Applying these conditions, 377 events have been selected as FCFV events. \
The expected number of background events from non beam-related sources in accidental
coincidence was estimated to be 0.0085.\\
From this point on, separate conditions have been applied to the $\nu_\mu$ and $\nu_e$ samples, allowing the selection of 120 $\nu_\mu$ and 28 $\nu_e$ candidates.\\

To identify $\nu_\mu$ CC  candidate  events  the following conditions have been applied: 

\begin{itemize}
\item There is only one reconstructed Cherenkov ring
\item The ring is $\mu$-like
\item The reconstructed momentum  is higher than 200  MeV
\item There are less than two reconstructed Michel electrons
\end{itemize}

The momentum cut (> 200 MeV) is applied to reject charged pions and misidentified electrons from the decay of un-observed muons and pions.  The requirement to have fewer than two Michel electrons rejects events with
additional unseen muons or pions.\\

For the selection of the  $\nu_e$ CC  candidate  events  the criteria listed below have been used: 

\begin{itemize}
\item There is only one reconstructed Cherenkov ring
\item The ring is e-like
\item The visible energy (Evis) is higher than 100  MeV
\item There is no reconstructed Michel electron
\item The reconstructed energy (Erec) is less than 1.25  GeV
\item The event is not consistent with a $\pi_0$  hypothesis 
\end{itemize}

The Evis requirement removes low energy neutral current (NC) interactions and electrons from the decay of unseen parents that are below Cherenkov threshold or fall outside the beam time window.  Since above 1.25 GeV
the intrinsic $\nu_e$ beam is dominant, a reconstructed energy below this threshold is also requested.\

Finally the same selection criteria have been applied to the corresponding Monte Carlo sample, obtaining the numbers of expected neutrino candidates for the no-oscillation hypothesis:
 they are 446 $\pm$ 23  for $\nu_\mu$ and 4.9 $\pm$ 0.6 for $\nu_e$ respectively. \\


\subsection{$\nu_\mu$ disappearance}

\begin{figure}[h!]
\begin{center}
\psfig{figure=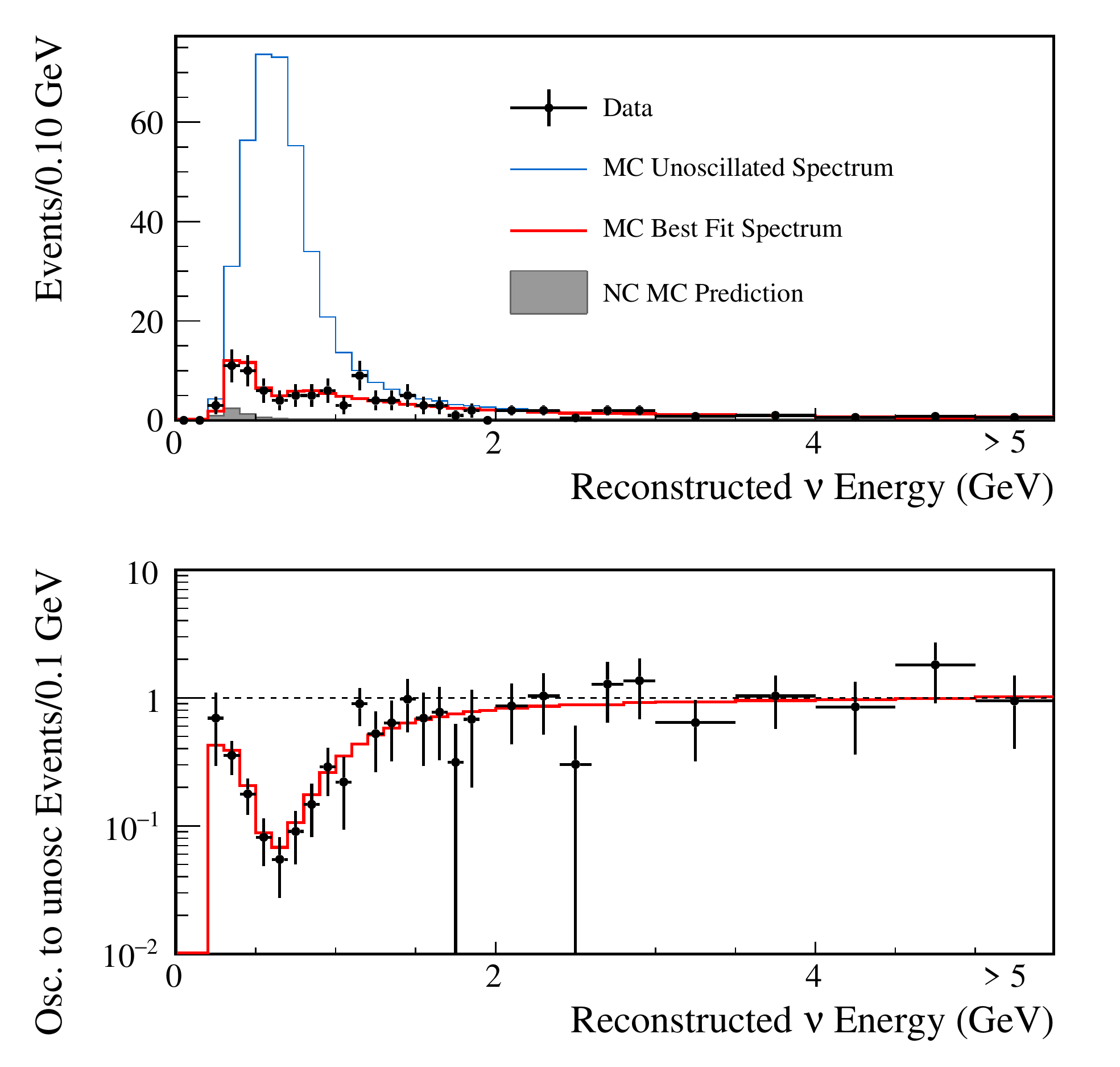,width=8cm}
\psfig{figure=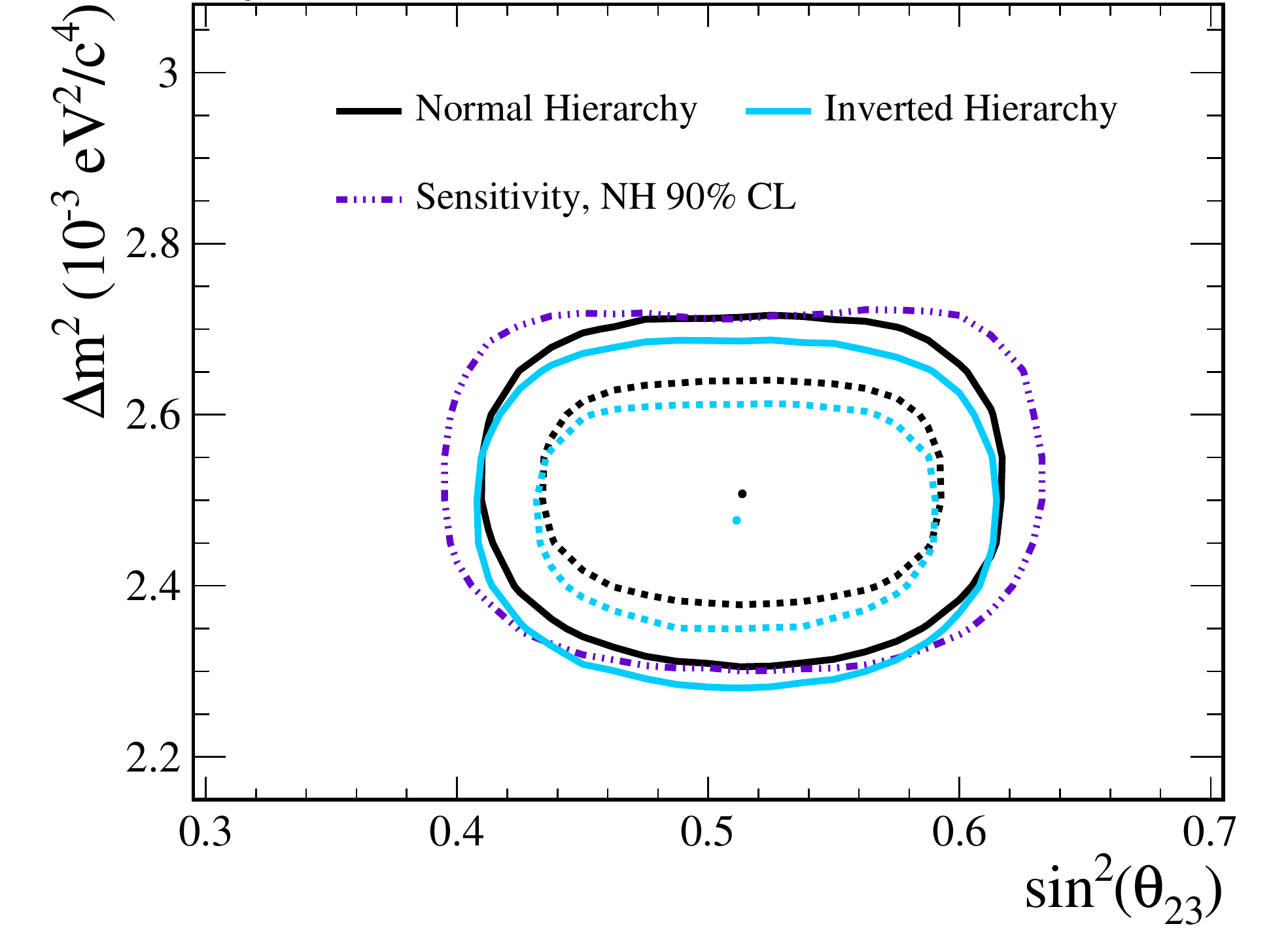,width=8cm}\caption{Right: the 68$\%$ (dashed) and 90$\%$ (solid) CL intervals for the  $\nu_\mu$-disappearance analysis assuming normal and inverted mass hierarchies. 
The 90$\%$ CL sensitivity contour for the normal hierarchy is overlaid for comparison.
Top(left): Reconstructed neutrino energy spectrum for data, best-fit prediction, and un-oscillated prediction. 
Bottom(left): Ratio of oscillated to un-oscillated events as a function of neutrino energy for the data and the best-fit spectrum.}
\label{fig:numu_dis}
\end{center}
\end{figure}

As reported in the previous subsection, 120 muon neutrino candidates have been observed in $6.57\times 10^{20}$ p.o.t. data, to be compared with the 446 $\pm$23 expected if no-oscillation is assumed. 
The neutrino energy distribution for the selected sample of 120 events is shown in Fig.\ref{fig:numu_dis}(left) together with the ratio of oscillated to un-oscillated events as a function of neutrino energy for the data and the best-fit spectrum. 
In Fig.\ref{fig:numu_dis}(left-top) is also visible the small contribution at low energy from NC as estimated by MC. The disappearance of muon neutrinos events as well as the distortion of the neutrino energy spectrum are evident.\\

The best-fit oscillation parameters measured under those conditions are: 
$\sin^2 ({\theta}_{23}) = 0.51\pm 0.056 $ and 
$|\Delta{m}^2_{32}| = 2.51 \pm 0.10 \times 10^{-3}$~eV$^2$ respectively \cite{12}.\\

The constraint in the two dimensional $\sin^2(\theta_{23})$ - $\Delta m^2_{32}$
plane for normal and inverted mass hierarchy is shown in Fig. \ref{fig:numu_dis}(right). The T2K results are consistent with those from SK\cite{29} and MINOS\cite{30}, and provide the most stringent constraint for $\sin^2(\theta_{23})$.\\

\subsection{$\nu_e$ appearance}

As presented in section 3.1, by analysing a data sample at SK corresponding to $6.57\times 10^{20}$ p.o.t. 
28 electron neutrino candidates have been observed, where 4.9$\pm$0.6 where expected from a  no-oscillation hypothesis. \
This result \cite{15} confirmed, with higher statistic, previous T2K claims \cite{13,14} based on  $1.43\times 10^{20}$ and $3.01\times 10^{20}$ p.o.t. respectively.\\
From a statistical point of view, the significance of the signal corresponds to 7.3 standard deviations. \
It can be concluded with certainty that for the first time the electron neutrino appearance has been observed in an almost pure  $\nu_\mu$ beam.\\ 
This result is very relevant  , in particular because it opens the possibility of new studies in the lepton sector of charge-parity (CP) violation, which provides a distinction in physical processes involving matter and antimatter. \\

\begin{figure}[h!]
\begin{center}
\psfig{figure=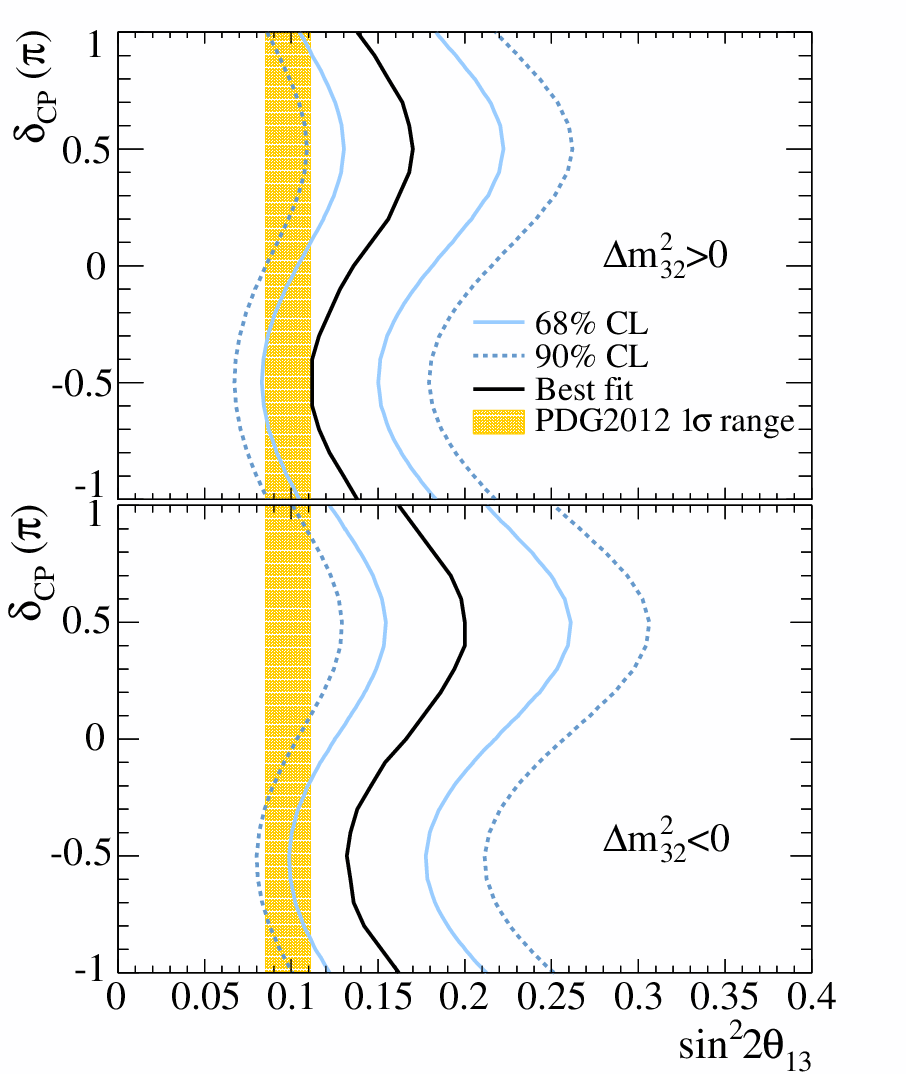,width=8cm}
\psfig{figure=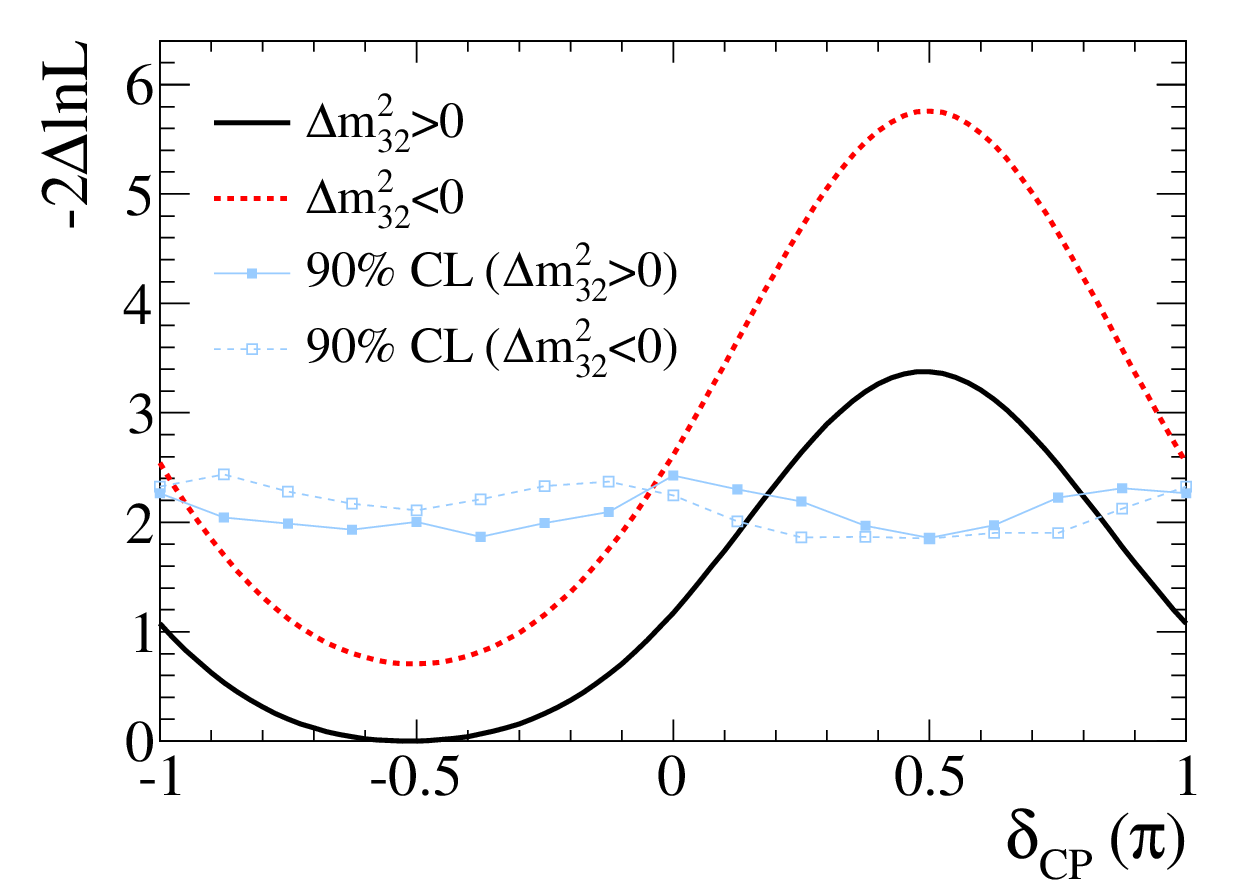,width=8cm}
\caption{Left: the 68\% and 90\% CL allowed regions for sin$^22\theta_{13}$, as a function of $\delta_{\mathrm{CP}}$ assuming normal hierarchy (top) and inverted hierarchy (bottom). 
The solid line represents the best fit sin$^2 2\theta_{13}$ value for given 
$\delta_{\mathrm{CP}}$ values. The values of sin$^2\theta_{23}$ and 
$\Delta m^{2}_{32}$ 
are varied in the fit with the constraint from \cite{12}. The shaded region shows the average $\theta_{13}$ value from the PDG2012 \cite{39}.
Right: The $-2\Delta \ln{\cal L}$ value as a function of $\delta_{\mathrm{CP}}$ for normal hierarchy (solid line) and inverted hierarchy (dotted line). 
The solid (dotted) line with markers corresponds to the 90\% CL limits for normal (inverted) hierarchy, evaluated by using the Feldman-Cousins method. 
The $\delta_{\mathrm{CP}}$ regions with values above the lines are excluded at 90\%~CL.}
\label{fig:nue_app}
\end{center}
\end{figure}

Constraints on oscillation parameters have been carefully calculated by comparing data and expectations. 
The allowed region in the  $sin^22\theta_{13}$- $\delta_{\mathrm{CP}}$ plane for normal mass hierarchy and
 inverted mass hierarchy are shown in Fig.\ref{fig:nue_app}(left) together with the constrains from reactor experiments \cite{15}. 
The overlap between T2K and the reactor results indicates that negative $\delta_{\mathrm{CP}}$  is favoured with a slight preference (67\%) for the normal mass hierarchy.\\

For $\delta_{\mathrm{CP}}$=0  and normal (inverted) hierarchy, the best-fit value with
a 68\% CL is $sin^22\theta_{13}$= 0.136 (+0.044/-0.033) ($sin^22\theta_{13}$=  0.166(+0.051/-0.042)).

At 90$\%$ confidence level and including reactor measurements, T2K excludes the region: 
$\delta_{\mathrm{CP}}$ =[0.15,0.83]$\pi$ for normal hierarchy and $\delta_{\mathrm{CP}}$=[-0.08,1.09]$\pi$ for inverted hierarchy. \\

The T2K and reactor data weakly favor the normal hierarchy with a Bayes Factor of 2.2.\\

The $-2\Delta \ln{\cal L}$ value as a function of $\delta_{\mathrm{CP}}$ for normal hierarchy (solid line) and inverted hierarchy (dotted line) is shown in Fig.\ref{fig:nue_app}(right)\cite{15}.
The likelihood is marginalized over sin$^22\theta_{13}$, sin$^2\theta_{23}$ and $\Delta m^{2}_{32}$.
The solid (dotted) line with markers corresponds to the 90\% CL limits for normal (inverted) hierarchy, evaluated by using the Feldman-Cousins method. 
The $\delta_{\mathrm{CP}}$ regions with values above the lines are excluded at 90\%~CL.\\

More recently \cite{32} the T2K collaboration published an analysis that for the first time combines measurements of muon neutrino disappearance and electron neutrino appearance to estimate four oscillation parameters and the mass hierarchy. 
\emph{Frequentist} and \emph{Bayesian} intervals have been used for combinations of these parameters, with and without the inclusion of recent reactor measurements.\\

\subsection{$\overline{\nu}_\mu$ disappearance (preliminary)}

Recently T2K reported \cite {25} an initial measurement of muon anti-neutrino disappearance using the accelerator-produced off-axis neutrino beam at JPARC.\\
The event selection applied at SK is unchanged with respect to to the neutrino beam mode previously described.\\
Using a dataset corresponding to $4.04 \times 10^{20}$ protons on target, $34$ fully contained $\mu$-like events were observed while $103$ events were predicted by Monte Carlo for the un-oscillated case. \\
The best-fit oscillation parameters measured under those conditions are: 
$\sin^2 (\overline{\theta}_{23}) = 0.45$ and 
$|\Delta\overline{m}^2_{32}| = 2.51 \times 10^{-3}$~eV$^2$ with 68\% confidence intervals of 0.38 - 0.64 and 2.26 - 2.80 ($\times 10^{-3}$~eV$^2$) respectively.\\

\begin{figure}[h!]
\begin{center}
\psfig{figure=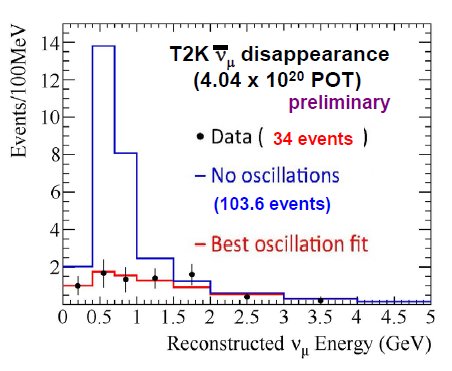,width=8cm}
\psfig{figure=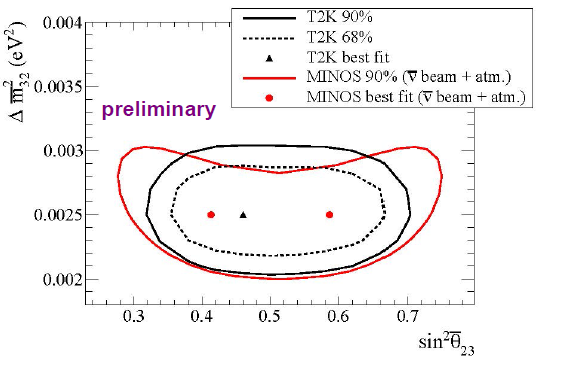,width=8cm}
\caption{Left: Distribution of reconstructed neutrino energy for 34 single ring $\mu$ like events observed in $4.04 \times 10^{20}$ p.o.t. anti-neutrino beam data. 
The expectations for no oscillation and for best fit oscillation
parameters are also shown. Right: Constraints on oscillation parameters $|\Delta\overline{m}^2_{32}|$ and $\sin^2 (\overline{\theta}_{23})$ obtained from
 $\overline{\nu}_\mu$ disappearance. Constraints from the MINOS \cite{30} experiment are also shown.}
\label{fig:anti_numu_dis}
\end{center}
\end{figure}

Preliminary results from the $\overline{\nu}_\mu$ disappearance analysis are illustrated in Fig.\ref{fig:anti_numu_dis}.\\

The distribution of reconstructed neutrino energy for the 34 single-ring $\mu$-like events, together with expectations for no oscillation hypothesis, are shown on the left side of the figure. The deficit is clearly seen.\\
On the right side the constraints on oscillation parameters $|\Delta\overline{m}^2_{32}|$ and $\sin^2 (\overline{\theta}_{23})$ obtained from this analysis are compared with previous results.\
Clearly they are in agreement with anti-neutrino measurements from both the MINOS and Super-Kamiokande collaborations, and also with precise measurements of neutrino disappearance from T2K.\\
In spite these results representing only 10$\%$ of the expected T2K anti-neutrino dataset, they are already competitive with both MINOS\cite{30} and SK\cite{31}, demonstrating the effectiveness of the off-axis beam technique.\\

\subsection{$\overline{\nu}_e$ appearance (preliminary)}

The selection process of  $\overline{\nu}_e$ candidates in SK is
exactly the same as for neutrino beam data. 
After all selections, three events remain as possible candidates of the 
 $\overline{\nu}_e$  appearance signal.
The expected number of background events are calculated by MonteCarlo assuming the absence of $\overline{\nu}_\mu$ - $\overline{\nu}_e$ oscillation.
Background events include ${\nu}_e$ appearance from 
 ${\nu}_\mu$ - ${\nu}_e$ oscillation,
misidentified ${\nu}_\mu$ (or $\overline{\nu}_\mu$), 
and original ${\nu}_e$ (or $\overline{\nu}_e$ )
from the decay of muons in the T2K beam line. 
The number of background events varies from 1.51 to 1.77, depending 
on mass hierarchy and $\delta_{\mathrm{CP}}$.\\
Obviously, the observation of three candidates is not significant evidence 
of $\overline{\nu}_e$ appearance but the collaboration plans to multiply 
by a factor 3 the statistic in the incoming years.\\

\section{T2K physics potential for $7.8 \times 10^{21}$ p.o.t.}
\label{sec:T2K_78}

\begin{figure}[h!]
\begin{center}

\psfig{figure=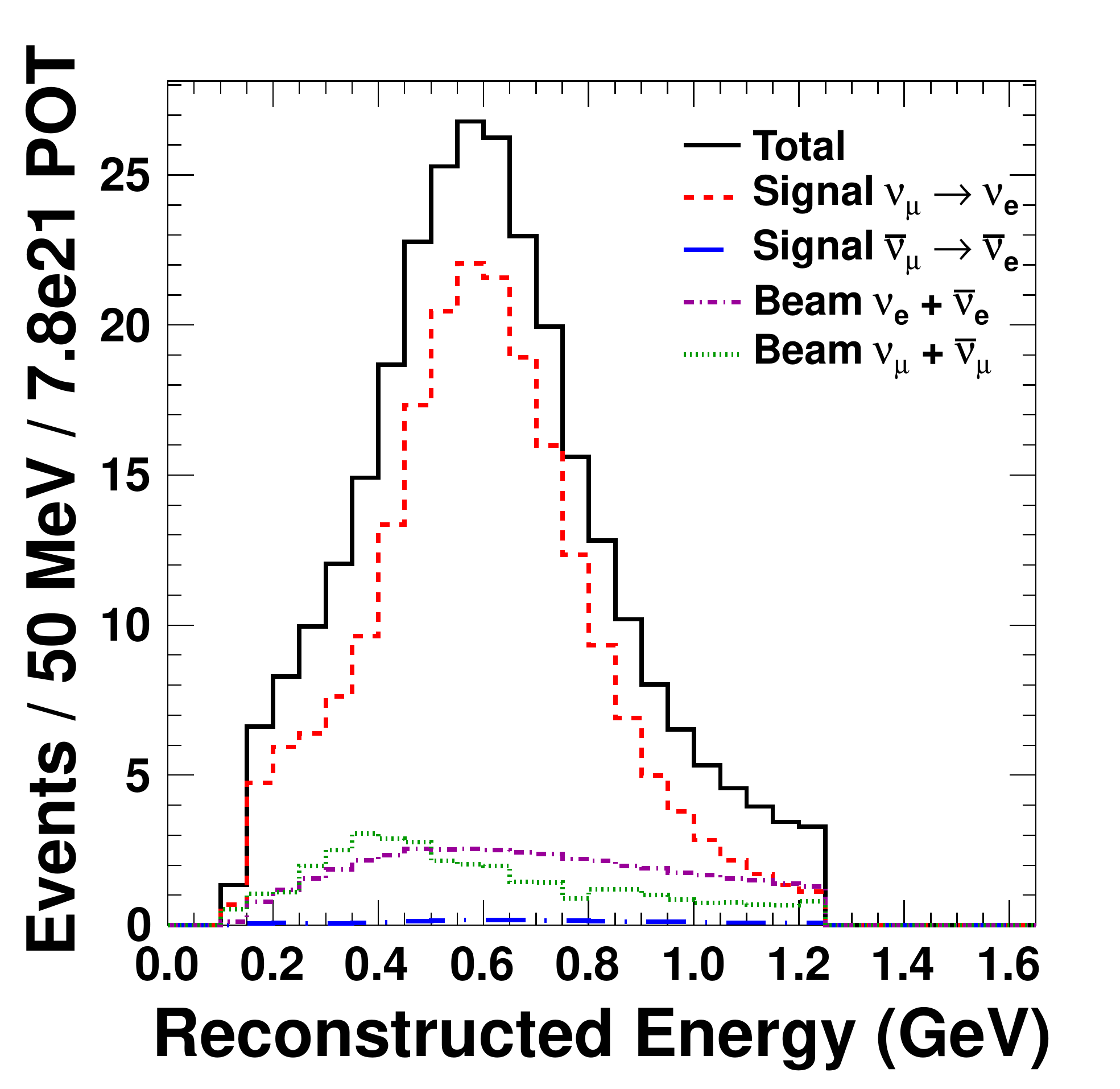,width=8cm}
\psfig{figure=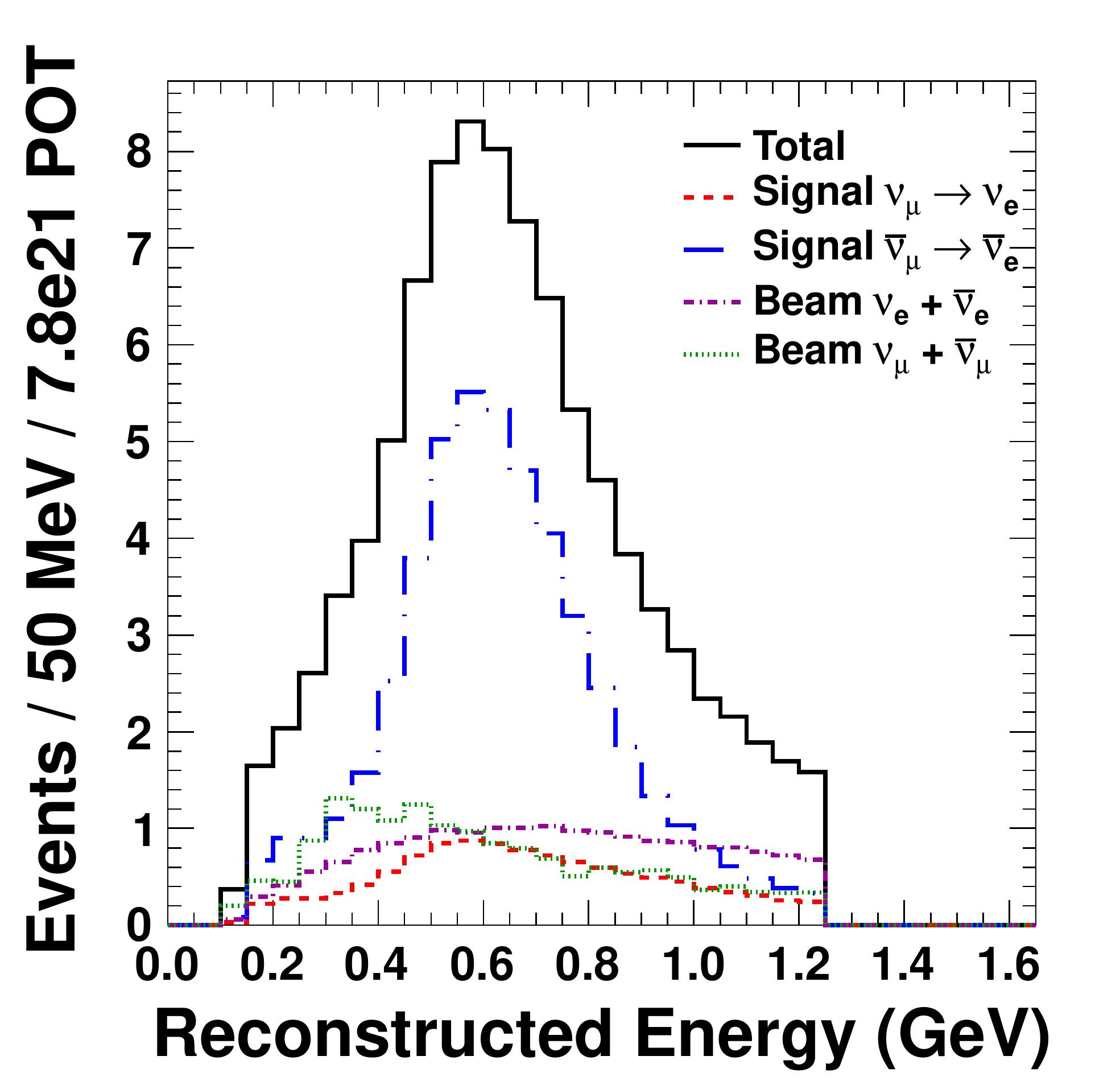,width=8cm}
\psfig{figure=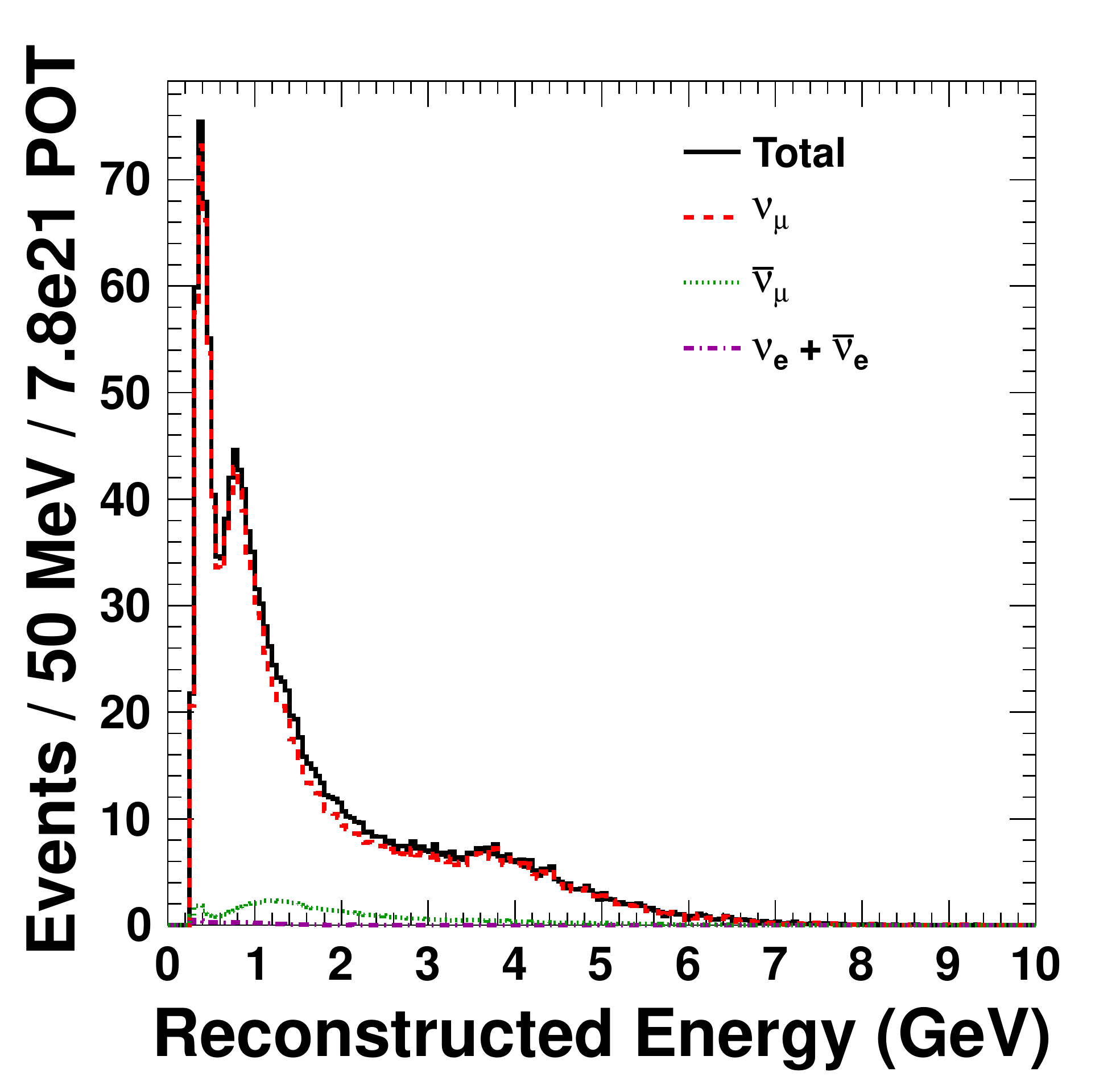,width=8cm}
\psfig{figure=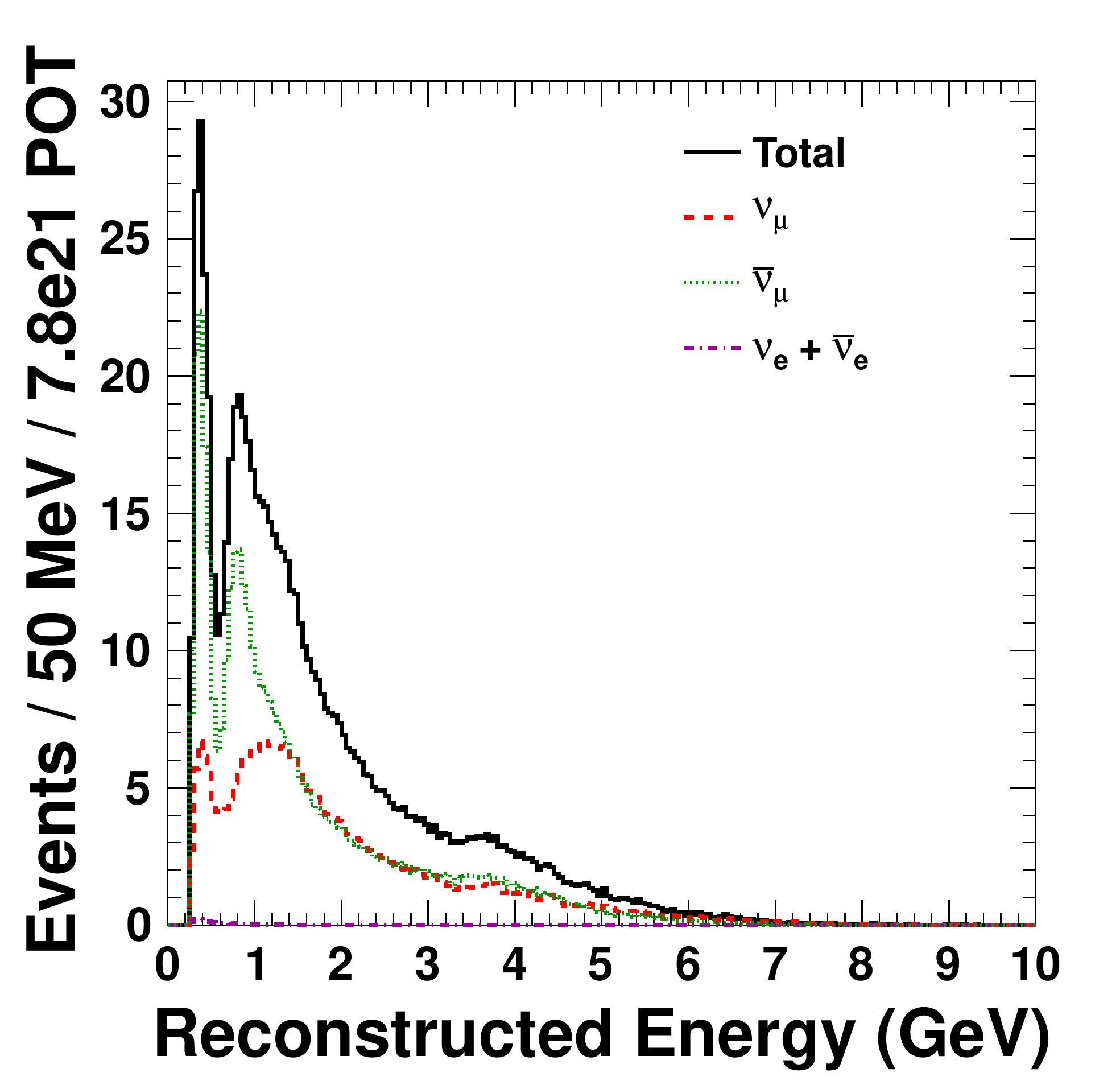,width=8cm}
\caption{
Top-left: \(\nu_e\) appearance reconstructed energy spectrum, 100\% \(\nu\)-mode running.
Top-right: \(\bar{\nu}_e\) appearance reconstructed energy spectrum, 100\% \(\bar{\nu}\)-mode running. 
Bottom-left: \(\nu_\mu\) disappearance reconstructed energy spectrum, 100\% \(\nu\)-mode running.
Bottom-right: \(\bar{\nu}_\mu\) disappearance reconstructed energy spectrum, 100\% \(\bar{\nu}\)-mode running.
All the oscillation parameters have been used in generating the spectra.\(\delta_{CP}\),
\(\sin^22\theta_{13}\), \(\sin^2\theta_{23}\), and \(\Delta m^2_{32}\) were left unknown in the fit, while \(\sin^22\theta_{12}\) and \(\Delta m^2_{21}\) are
assumed fixed to the values given in Table \ref{tab:truepars}} 

\label{fig:RecEspect}
\end{center}
\end{figure}

The observation of the electron neutrino appearance in a muon neutrino beam by T2K and 
the high-precision measurement of the mixing angle $\theta_{13}$  by reactor experiments
have led to a re-evaluation of the physics potential 
of T2K for the approved data set ($7.8 \times 10^{21}$ p.o.t.) that, according to the latest plans of the accelerator group in JPARC, will be achieved by $\sim $2020.\\
In particular in \cite{33} the sensitivities for CP violation in neutrinos,
non-maximal $\sin^22\theta_{23}$, octant of $\theta_{23}$,
and mass hierarchy have been explored for T2K alone and in combination with NO$\nu$A and reactor experiments results.\\
Special care was also taken in studying the effect coming for various combinations of
$\nu$-mode and \(\bar{\nu}\)-mode data-taking. 
In fact the probability of \numu{}\goesto{}\nue{} is slightly different from the
probability for $\bar{\nu}_\mu$\goesto{}$\bar{\nu}_e$ oscillation because of the swap 
in the sign of the $\delta_{CP}$ term. \
This implies that  the anti-neutrino oscillation
probability is larger (smaller) than the neutrino oscillation probability for positive (negative) $\delta_{CP}$ by up to 25\%.\
Accordingly, comparison of oscillation probabilities between neutrino-mode 
and anti-neutrinos-mode could help in the determination of  the $\delta_{CP}$ value.\\

\subsection{T2K alone}
A three-flavor analysis combining appearance and disappearance, 
for both \(\nu\)-mode, and \(\bar{\nu}\)-mode 
have been performed assuming the expected full statistics 
of $7.8\times10^{21}$ p.o.t. .\

The selection of candidate events in SK was done using the same criteria described in Sec 3.1.\\
The study includes either statistical errors only or 
statistical and systematic errors established for the 2012 oscillation analyses.  
In addition, signal efficiency and background are taken into account. \\
It should be emphasized that this evaluation is conservative, considering that the analyses performed 
on data collected in the years 2013-2015 already showed an improved precision by about 20\%, or more, with respect to the numbers quoted in \cite{33}.\\

Reconstructed appearance and disappearance energy spectra generated 
for the approved full T2K statistics, assuming a data-taking condition
of either 100\% \(\nu\)-mode or 100\% \(\bar{\nu}\)-mode, are shown in Fig.\ \ref{fig:RecEspect}.\ 
All the oscillation parameters have been used in generating the spectra.\(\delta_{CP}\),
\(\sin^22\theta_{13}\), \(\sin^2\theta_{23}\), and \(\Delta m^2_{32}\) were left unknown in the fit, while \(\sin^22\theta_{12}\) and \(\Delta m^2_{21}\) are
assumed fixed to the values given in Table \ref{tab:truepars}. 
The expected number of  $\nu_e$ or $\bar{\nu}_e$ appearance events 
at $7.8\times10^{21}$~p.o.t. obtained from \cite{33} are shown  in Table \ref{tab:nevapp} for two different values of 
$\delta_{CP}$ (-90\(\degree\),0\(\degree\)).\
The number of events is broken down into those coming from appearance
signal or intrinsic beam background events that undergo charged current (CC)
interactions in SK, or beam background events that undergo neutral current (NC) interactions.\\
The values given in Table \ref{tab:nevapp} indicate that statistics is much more favourable for the \(\nu\)-mode, where a signal 3 times larger is expected. However, only combining \(\nu\)-mode
and  \(\bar{\nu}\)-mode will give an advantage in constraining $\delta_{CP}$.\\

\begin{table}[htbp]
\caption[Nominal Oscillation Parameter Values]{Nominal values of the oscillation parameters. }
\begin{center}
\begin{tabular}{  l || c | c | c | c | c | c | c } \hline
Parameter & \(\sin^22\theta_{13}\) & \(\delta_{CP}\) & \(\sin^2\theta_{23}\) &
\(\Delta m^2_{32}\) &
Hierarchy & \(\sin^22\theta_{12}\) & \(\Delta m^2_{21}\)\\ \hline
Nominal & 0.1 & 0 & 0.5 & \(2.4\times10^{-3}\) & normal & 0.8704 &
\(7.6\times10^{-5}\) \\
Value & &  &  &  eV\(^2\) & & &  eV\(^2\) \\ \hline
\end{tabular}
\end{center}  
\label{tab:truepars} \end{table}

\begin{table}[htbp] 
\caption[Number of $\nu$ Appearance Events]{Expected numbers of 
$\nu_e$ or $\bar{\nu}_e$ appearance events at $7.8\times10^{21}$~p.o.t..  The
number of events is broken down into those coming from: appearance
signal or intrinsic beam background events that undergo charged current (CC)
interactions in SK, or beam background events that undergo neutral current (NC) interactions.}
\begin{center}

\begin{tabular}{  c | c | c | c | c | c | c | c  } \hline 
& & & Signal & Signal & Beam CC & Beam CC & \\
& \(\delta_{CP}\) & Total & \(\nu_{\mu} \rightarrow \nu_e\) & \(\bar{\nu}_{\mu} \rightarrow
\bar{\nu}_e\) & \(\nu_e + \bar{\nu}_e \) & \(\nu_{\mu} + \bar{\nu}_{\mu} \) &
NC\\ \hline \hline
100\% $\nu$-mode  & 0\(\degree\) & 291.5 & 211.9 & 2.4 & \multirow{2}{*}{41.3} & \multirow{2}{*}{1.4} & \multirow{2}{*}{34.5} \\ 
100\% $\nu$-mode  & -90\(\degree\) & 341.8 & 262.9 & 1.7 & & & \\ \hline 
100\% $\bar{\nu}$-mode & 0\(\degree\) & 94.9 & 11.2 &  48.8 & \multirow{2}{*}{17.2} & \multirow{2}{*}{0.4} & \multirow{2}{*}{17.3} \\ 
100\% $\bar{\nu}$-mode & -90\(\degree\) & 82.9 & 13.1 & 34.9 &  &  &  \\ \hline 
\end{tabular}
\end{center} 
\label{tab:nevapp} \end{table}

\begin{figure}[h!]
\begin{center}

\psfig{figure=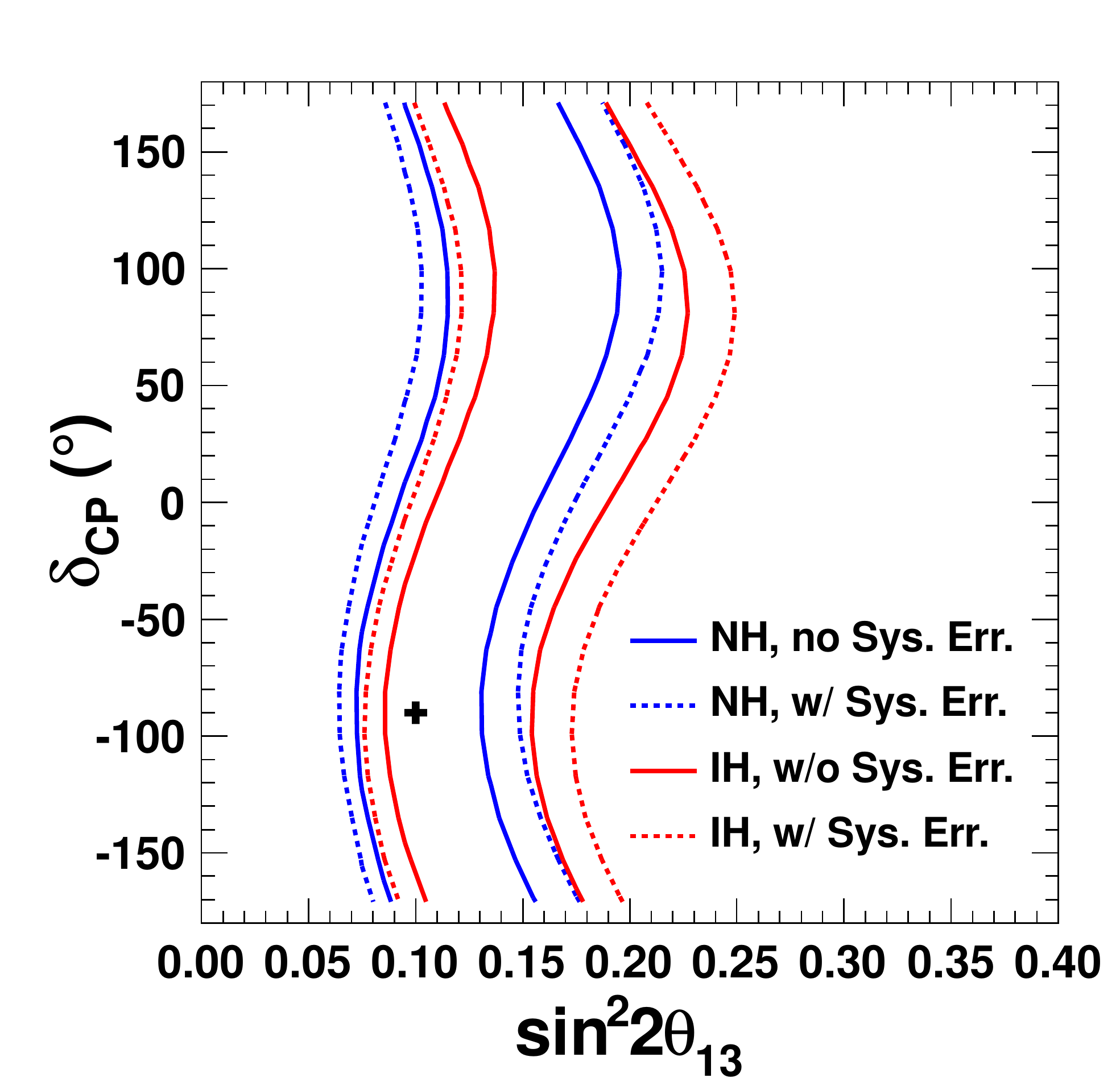,width=8cm}
\psfig{figure=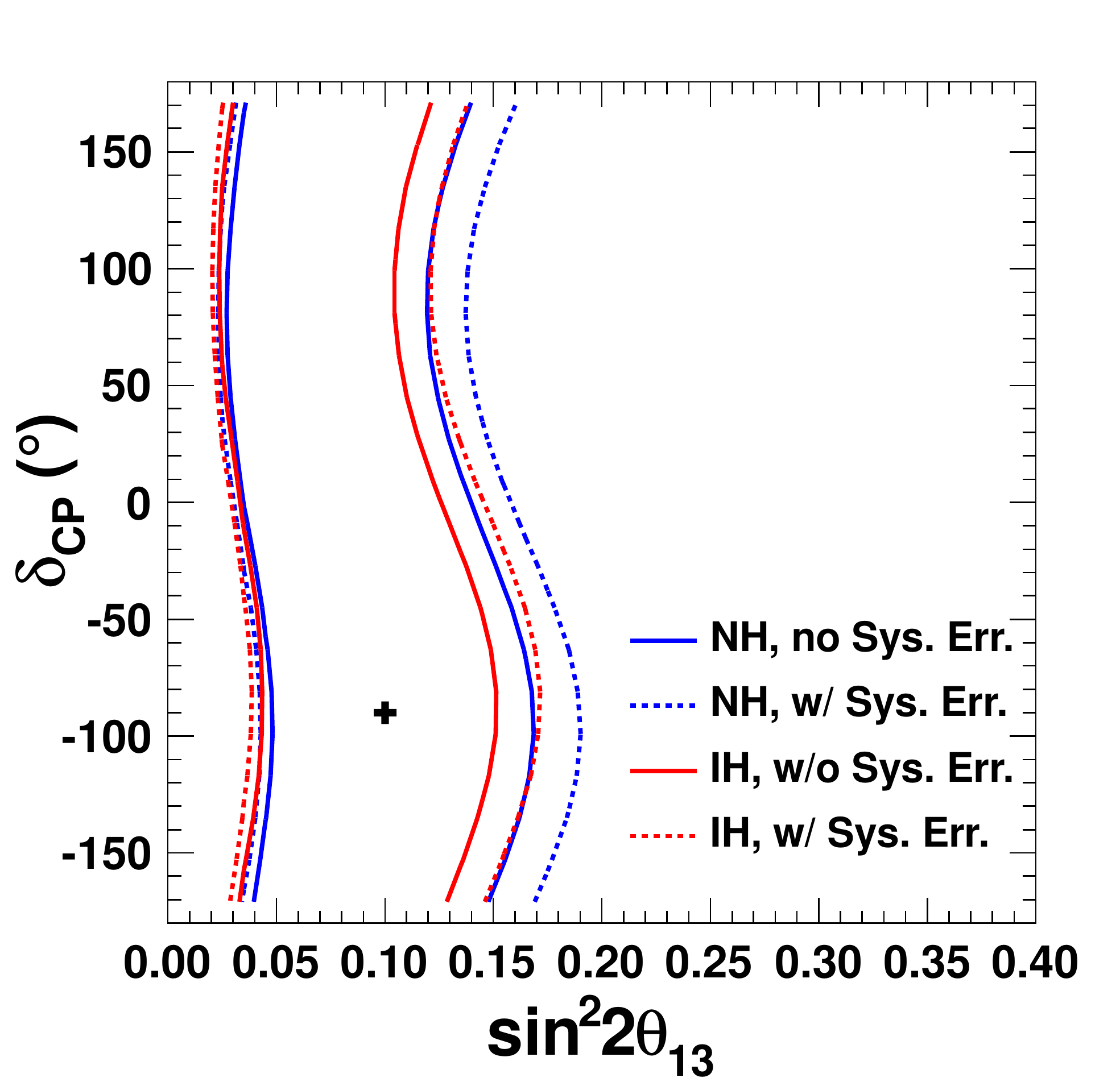,width=8cm}
\psfig{figure=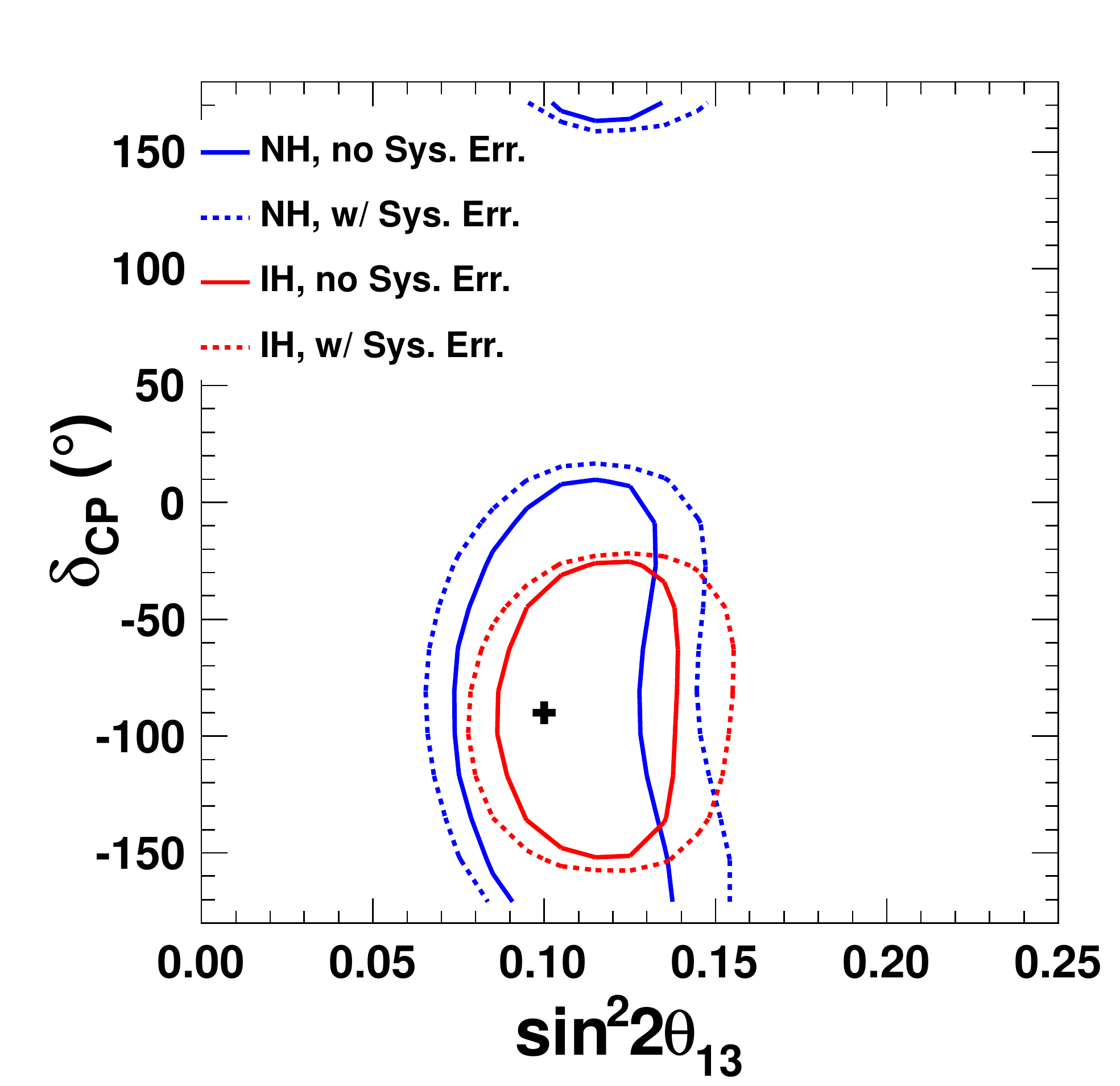,width=8cm}
\caption{Expected \(\delta_{CP}\) vs\ \(\sin^22\theta_{13}\) 90\% C.L.\ 
intervals, where (top-left) and (top-right) figures are given for neutrino and anti-neutrino mode running for 50\% of the full T2K p.o.t. each, and the bottom figure demonstrate the sensitivity of the 
total T2K p.o.t. with 50\% \(\nu\)-mode plus 50\% \(\bar{\nu}\)-mode running.  
Contours are plotted for the case of true \(\delta_{CP} = -90\degree\) and NH. The blue curves are fits  assuming the correct MH(NH), while the red are fits  assuming the incorrect MH(IH), and contours are plotted from the minimum \(\chi^2\)value for both MH assumptions.
The solid contours are with statistical error only, while the dashed contours 
include the systematic errors used in the 2012 oscillation analysis
assuming full correlation between \(\nu\)- and \(\bar{\nu}\)-mode running errors.
} 

\label{fig:nu-nubar}
\end{center}
\end{figure}

This effect is clearly illustrated  in Figure~\ref{fig:nu-nubar}.\\
In this example, top-left and top-right panels of  Figure~\ref{fig:nu-nubar}  show 
the \(\delta_{CP}\) vs\ \(\sin^22\theta_{13}\)  90\% C.L.\ intervals,
each given for 50\% of the full T2K p.o.t., of \(\nu\) and \(\bar{\nu}\)-mode 
at true \(\delta_{CP} =-90\degree\) assuming NH 
\footnote{if the fit is assuming the correct Mass Hierarchy (MH) it is called NH, while if it is assuming the incorrect MH is called IH} and without a reactor constraint.\\
When the two contours are combined in Fig.\ \ref{fig:nu-nubar} (bottom), it becomes evident that $\delta_{CP}$ can be constrained without any requirement from external data.\\

\begin{figure}[h!]
\begin{center}

\psfig{figure=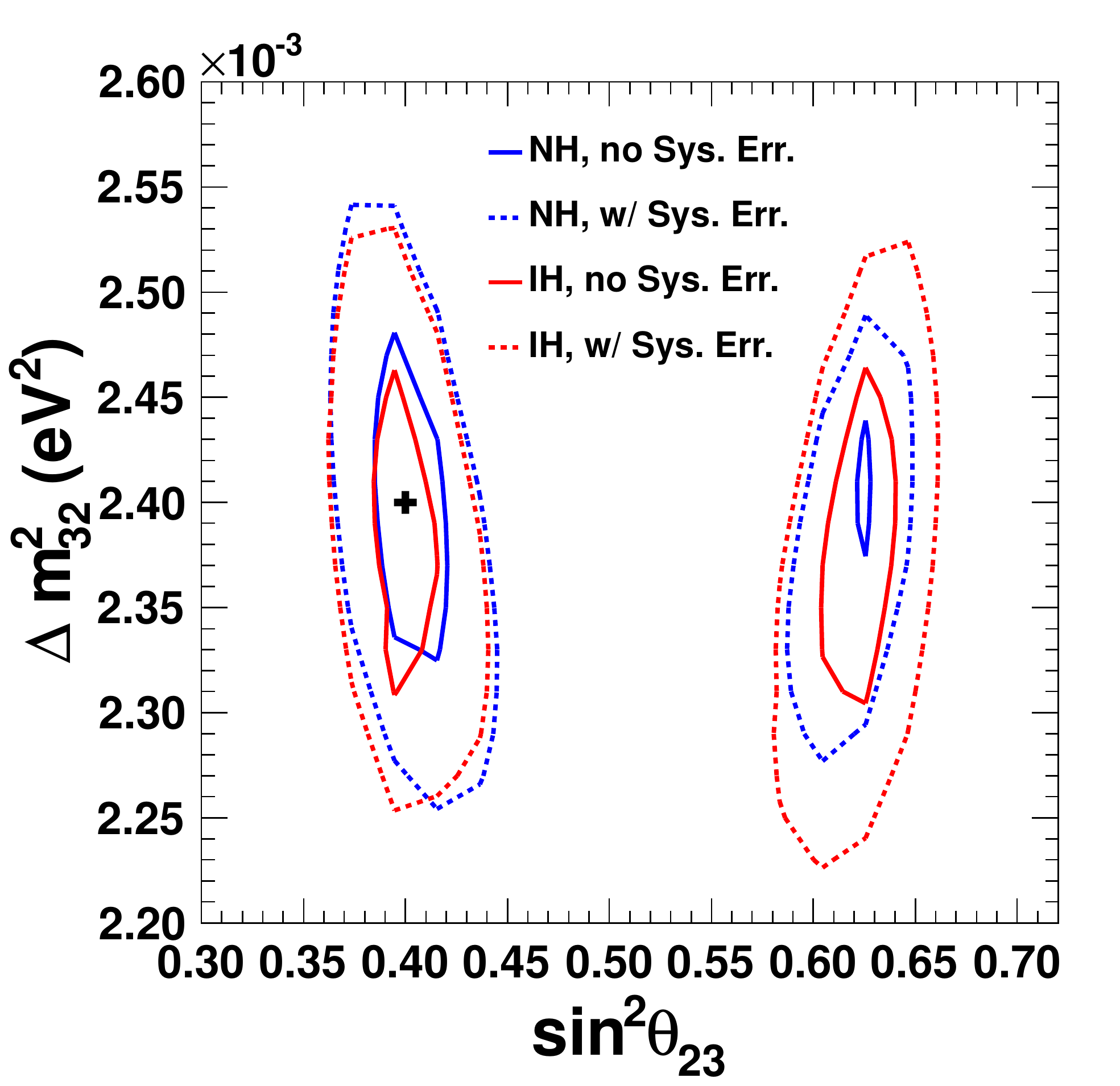,width=8cm}
\psfig{figure=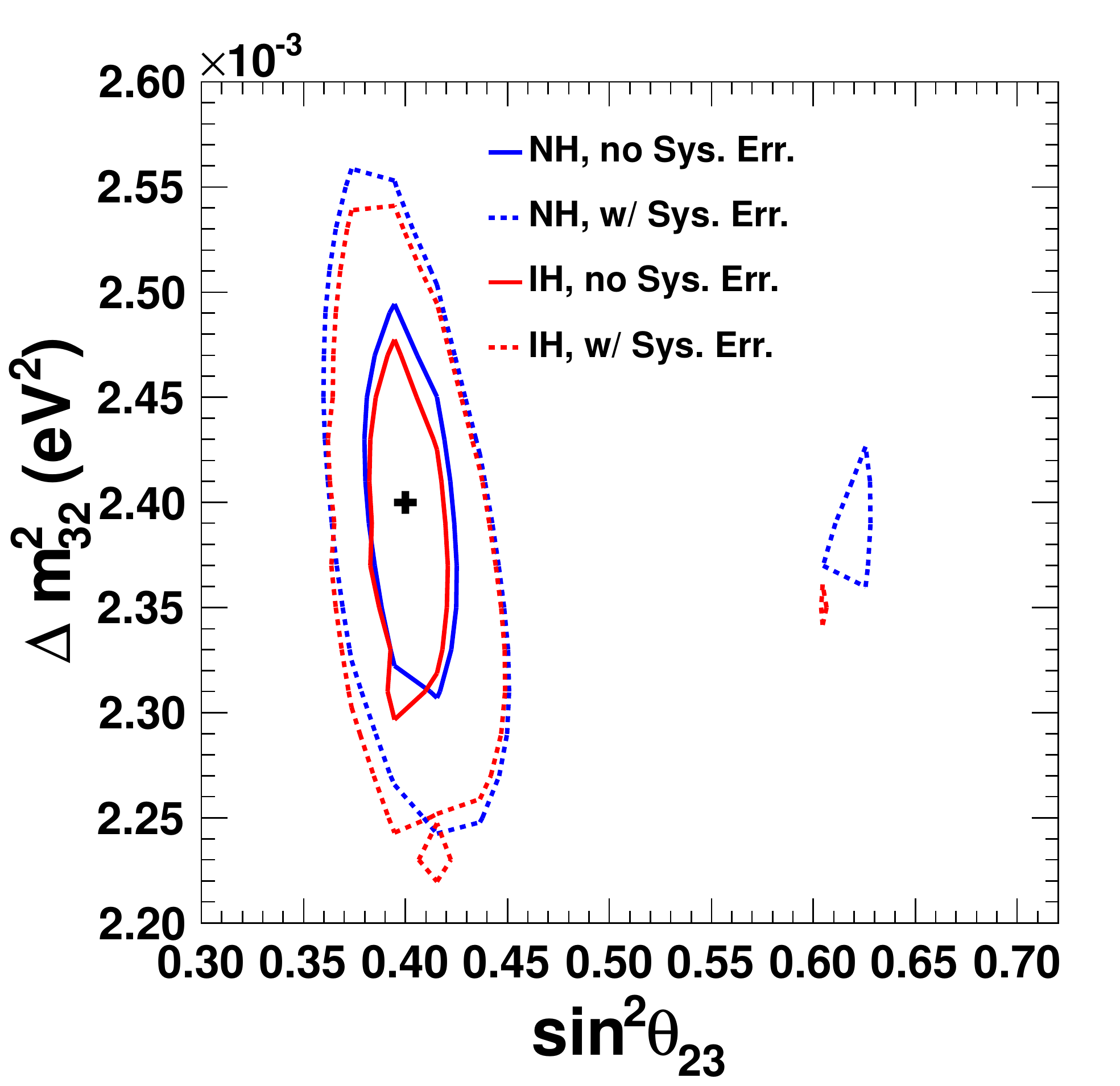,width=8cm}
\caption{
\(\Delta m^2_{32}\) vs\ \(\sin^2\theta_{23}\) 90\% C.L.\ intervals for
 \(7.8\times10^{21}\)~p.o.t. (50\% \(\nu\)-mode plus 50\% \(\bar{\nu}\)-mode running).  
 Contours are plotted for the case of true \(\delta_{CP} = 0\degree\), 
\(\sin^2\theta_{23}=0.4\), $\Delta m^2_{32} = 2.4\times10^{-3}~\mbox{eV}^2$ and NH. 
The plot on the left does not include the reactor constraint; the plot on the right includes it.
The blue curves show fits assuming the correct MH(NH), while the red ones show fits assuming the incorrect MH(IH). The solid contours are with statistical error only, while the dashed contours
include the 2012 systematic errors fully correlated between \(\nu\)- and
\(\bar{\nu}\)-mode.} 

\label{fig:theta23}
\end{center}
\end{figure}

However, for other parameters (like $\theta_{23}$) the constraint from the reactor measurements
is very important and it is not avoidable if one wishes to discriminate the octant and cancel the degeneracies.\\ 
Figure~\ref{fig:theta23} shows, as an example, the 90\% C.L.\ 
regions for $\Delta m_{32}^2$ vs\ $\sin^2{\theta_{23}}$
at the full T2K statistics  for $\sin^2{\theta_{23}}=0.4$. 
In this case the  $\theta_{23}$ octant cannot be resolved  only by combining both $\nu$-mode and $\bar{\nu}$-mode data (Fig.~\ref{fig:theta23}-left).\
In fact the reactor constraint on $\theta_{13}$ should be included 
 to resolve degeneracies between the oscillation parameters
 \(\sin^2\theta_{23}\), \(\sin^22\theta_{13}\), and \(\delta_{CP}\) 
 as clearly illustrated in Fig.~\ref{fig:theta23}-right.\\

\newpage

\subsection{T2K + NO$\nu$A}
 
\begin{figure}[h!]
\begin{center}
\psfig{figure=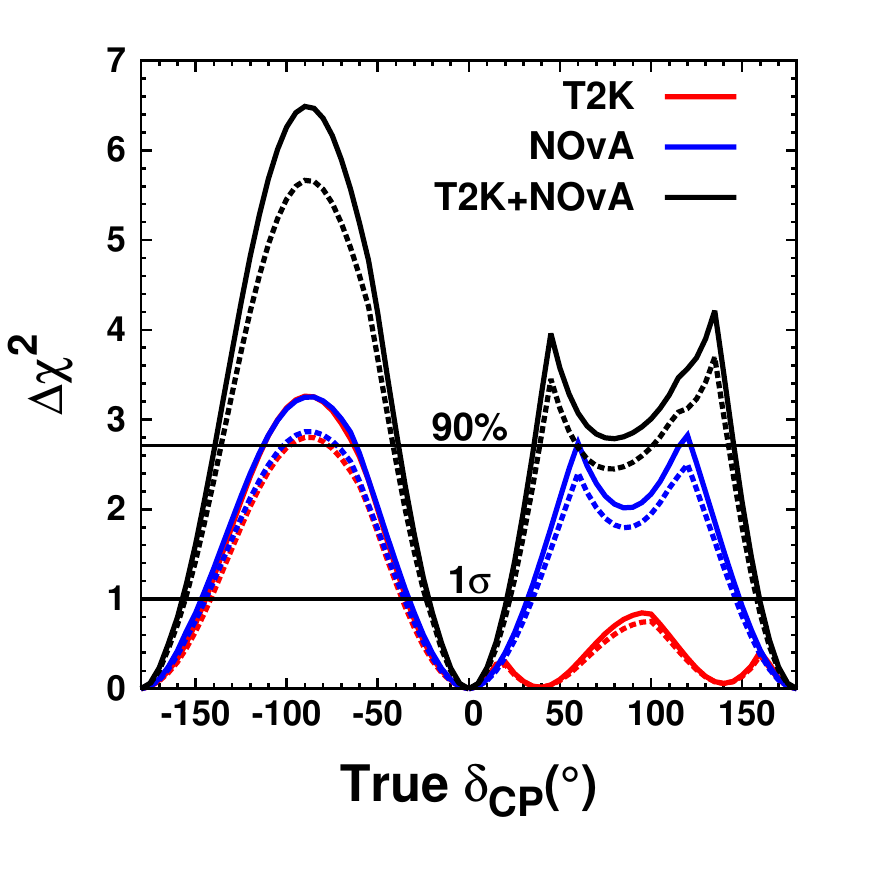,width=8cm}
\psfig{figure=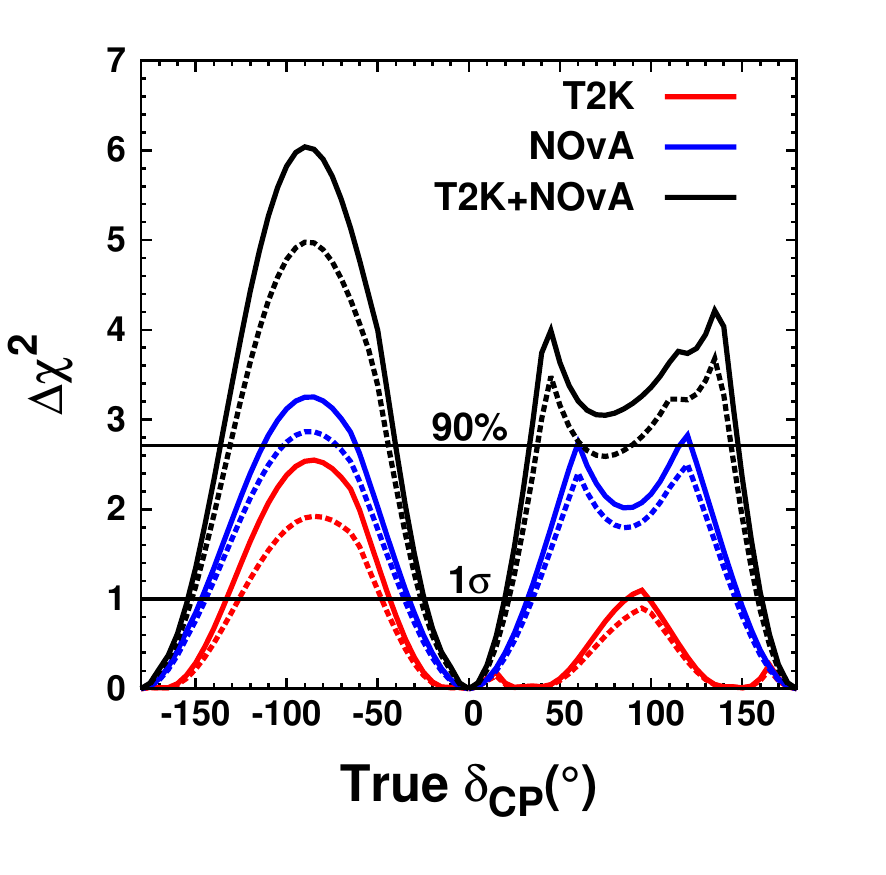,width=8cm}
\caption{The predicted $\Delta\chi^2$ for rejecting the
$\sin\delta_{CP}=0$ hypothesis, 
as a function of \(\delta_{CP}\) for T2K~(red), NO$\nu$A~(blue), 
and T2K+NO$\nu$A~(black), assuming NH. Dashed (solid) curves indicate studies 
where normalization systematics are (not) considered.\
Left:1:0 T2K, 1:1 NO$\nu$A $\nu$:$\bar{\nu}$.
Right:1:1 T2K, 1:1 NO$\nu$A $\nu$:$\bar{\nu}$} 
\label{fig:t2K+Nova-NH}
\end{center}
\end{figure}

\begin{figure}[h!]
\begin{center}

\psfig{figure=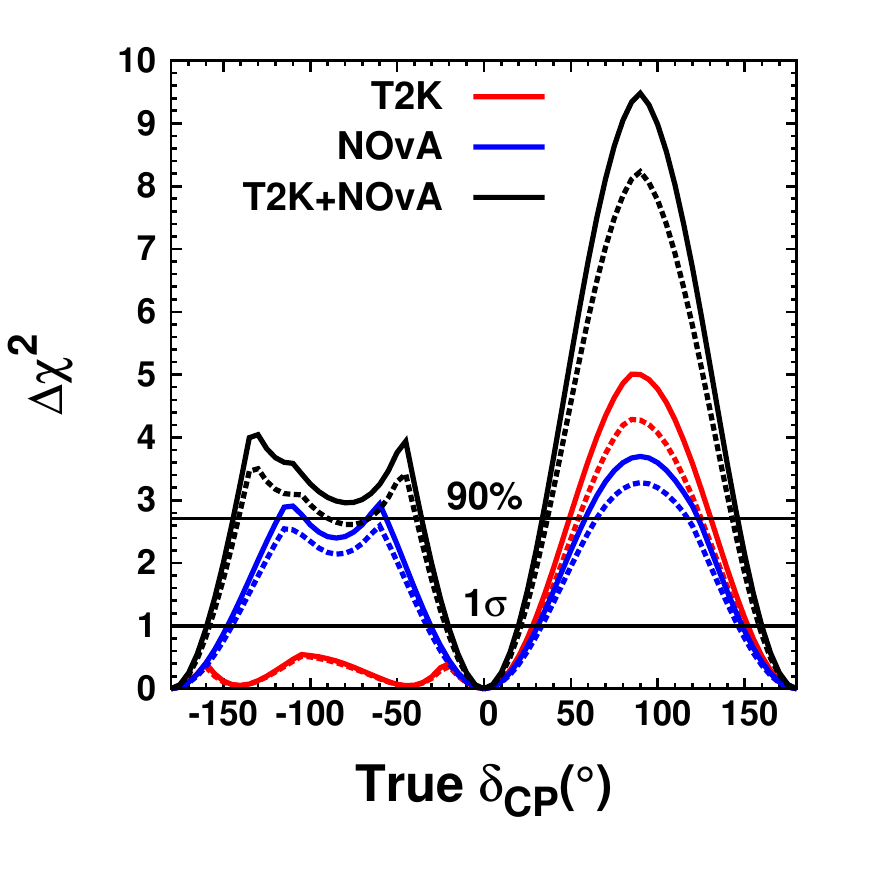,width=8cm}
\psfig{figure=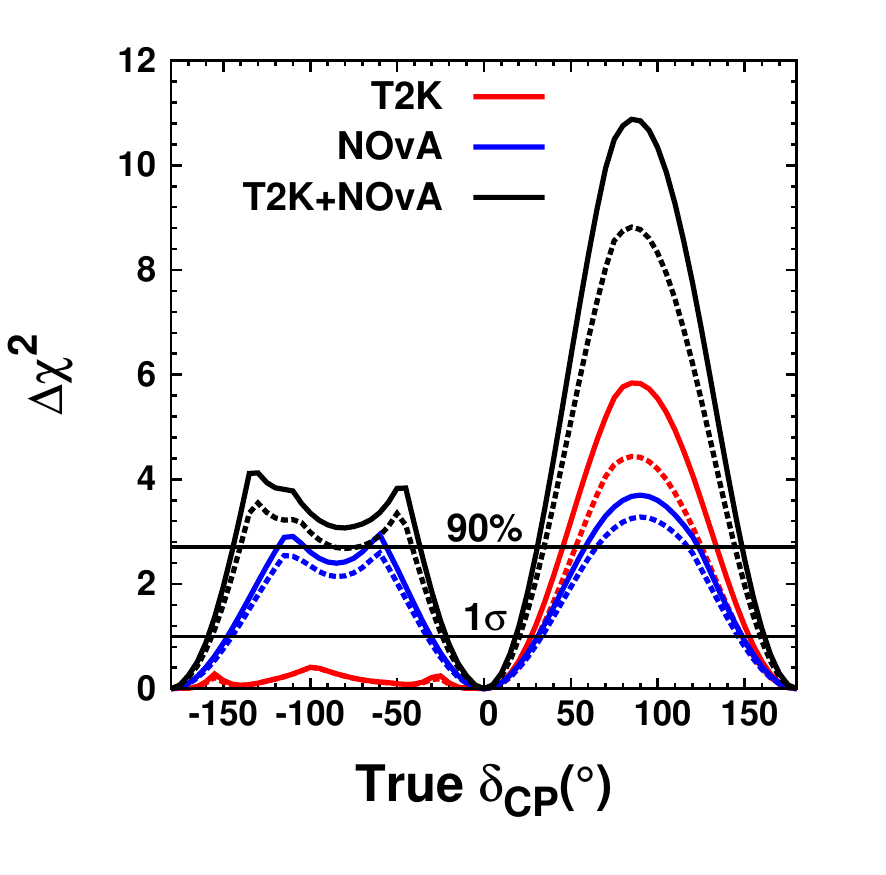,width=8cm}

\caption{The predicted $\Delta\chi^2$ for rejecting the
$\sin\delta_{CP}=0$ hypothesis, 
as a function of \(\delta_{CP}\) for T2K~(red), NO$\nu$A~(blue), 
and T2K+NO$\nu$A~(black), assuming IH. Dashed (solid) curves indicate studies 
where normalization systematics are (not) considered.\
Left:1:0 T2K, 1:1 NO$\nu$A $\nu$:$\bar{\nu}$.
Right:1:1 T2K, 1:1 NO$\nu$A $\nu$:$\bar{\nu}$} 

\label{fig:t2K+Nova-IH}
\end{center}
\end{figure}

Since the ability of T2K to measure the value of $\delta_{\rm{CP}}$ 
is greatly enhanced by the knowledge of the mass hierarchy, the same study \cite{33} has also incorporated the impact of expected data from the NO$\nu$A experiment using the GLoBES tools\cite{40}.\\
The NO$\nu$A experiment~\cite{21}, which started operating in 2014,
has a longer baseline ($810$ km) and higher peak neutrino energy 
($\sim2$ GeV) than T2K. Accordingly,  the impact of
the matter effect on the predicted far detector event spectra is larger 
in  NO$\nu$A ($\sim 30\%$) than in T2K ($\sim 10\%$),
leading to a better sensitivity to the mass hierarchy.\\ 
The study assumes the T2K final statistic ($7.8 \times 10^{21}$ p.o.t.) \cite{19}and $1.8 \times 10^{21}$ p.o.t. in $\nu$-mode and $1.8 \times 10^{21}$ p.o.t. \(\bar{\nu}\) for  NO$\nu$A \cite{21}.\
The result is illustrated in Figs.(\ref{fig:t2K+Nova-NH},\ref{fig:t2K+Nova-IH}) for the
NH and IH case respectively. The plots on the left assume a data-taking condition
of 100\% \(\nu\)-mode for T2K and  50\% \(\nu\) 50\% \(\bar{\nu}\)-mode for NO$\nu$A. 
The plots on the right assume a data-taking condition
of 50\% \(\nu\) 50\% \(\bar{\nu}\)-mode for both T2K and NO$\nu$A. \

Because of the complementary nature of these two experiments, when T2K data is combined with data from NO$\nu$A,
the region of oscillation parameter space 
where there is sensitivity to observe a non-zero $\delta_{CP}$ 
is substantially increased compared to when each experiment is analyzed alone.\\
The results of the studies in \cite{33} are actually used to guide the optimization of the future run plan for T2K.\\

\newpage
\section{T2K physics potential for $20 \times 10^{21}$ p.o.t. or more}
\label{sec:T2K_250}

In summer 2015 the T2K Collaboration has been considering an extension of the data taking run beyond the approved total $7.8 \times 10^{21}$  to $20\times 10^{21}$ p.o.t. or more ($25\times 10^{21}$ p.o.t.)
In fact, from studies \cite{34,35}, based on the same considerations described in Sec. 4.1, 
an enhancement of the statistics by a factor 3 or more  could possibly lead to a 3 $\sigma$
measurement excluding  $\sin(\delta_{CP})= 0$ (depending on the true value 
of  $\delta_{CP}$ and on the knowledge of the MH) and showing the first evidence of CP violation in the lepton sector.\\
In this section we will give a short summary of those recent studies with the \emph{caveat} 
that all the plots showed below must be considered \emph{work in progress}.\\ 

\subsection{JPARC Beam Update}

\begin{figure}[h!]
\begin{center}
\psfig{figure=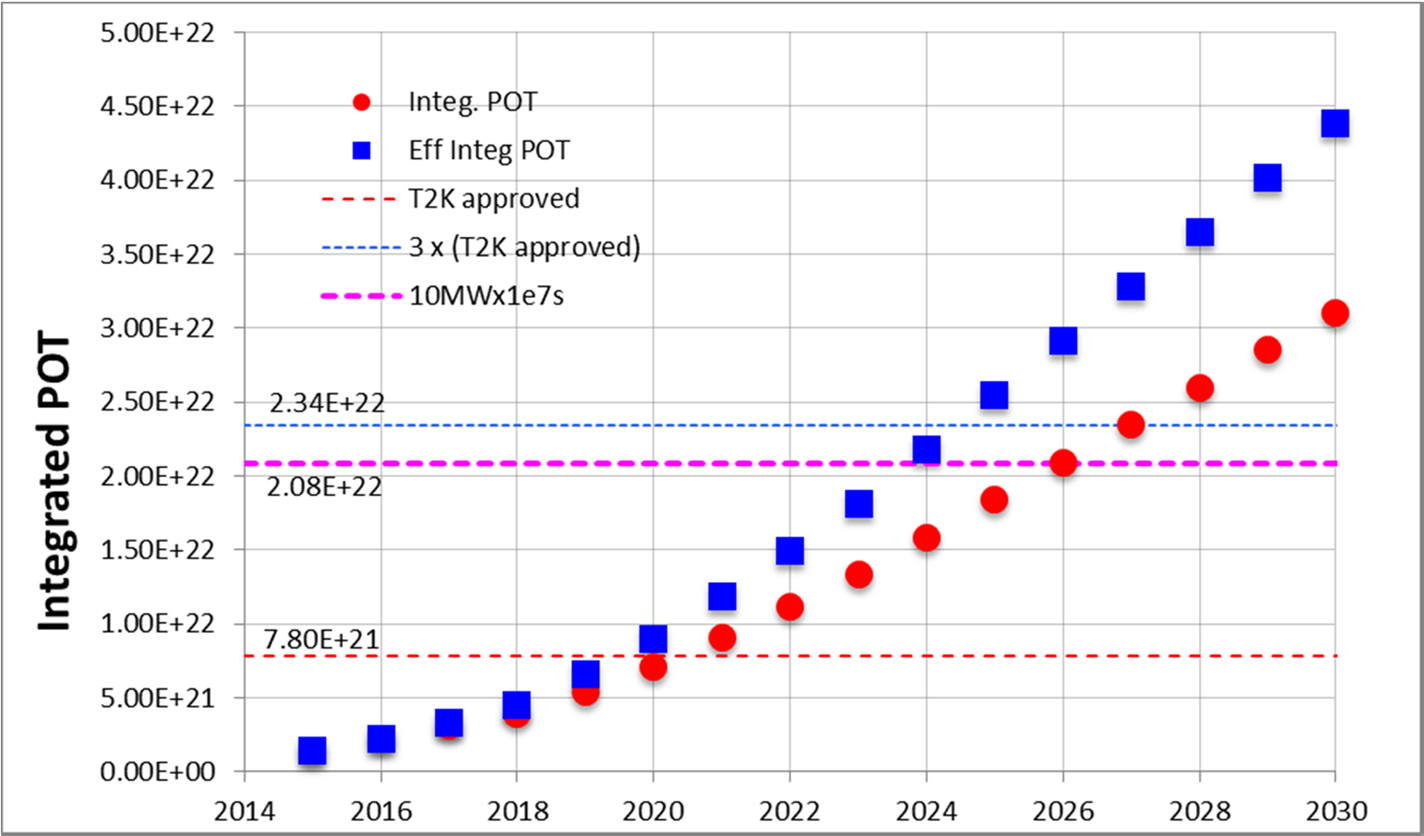,width=10cm}
\caption{The extended T2K p.o.t. projection, based on the last JPARC beam schedule (red dots) together with a  new possible J-PARC beam upgraded schedule as envisaged in \cite{36} (blue dots). The new projection assumes an effective
p.o.t., calculated including additional hardware upgrades and analysis improvements in SK}
\label{fig:IntPOT}
\end{center}
\end{figure}

The extended T2K p.o.t. projection, based on the latest JPARC beam schedule (red dots), is shown in Fig.\ref{fig:IntPOT}  together with a  new possible J-PARC beam upgraded schedule as 
envisaged in \cite{36} (blue dots). The extended T2K p.o.t. projection includes the MR (Main Ring) update 
(recently approved) that will allow to reach a power of 750KW in 2019 and up to 1.3MW when the repetition cycle will be reduced from 2.4s to 1.3s.\

\begin{figure}[h!]
\begin{center}
\psfig{figure=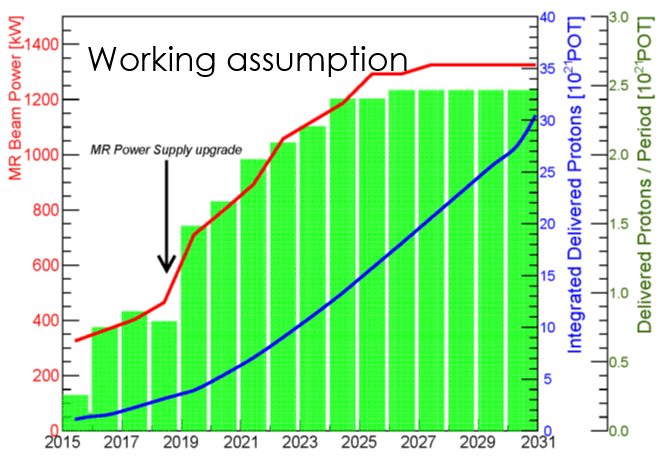,width=10cm}
\caption{Anticipated MR beam power and POT accumulation vs. calendar year.\cite{38}}
\label{fig:MR-Power}
\end{center}
\end{figure}

The MR beam power time evolution together with the possible data accumulation is shown 
in Fig.\ref{fig:MR-Power}  \cite{38}, where 5 months neutrino beam operation each year and realistic running time efficiency are assumed. \\

The new projection shown in Fig.\ref{fig:IntPOT} (blue dots) assumes an effective p.o.t. calculated by also taking into account  additional hardware upgrades and some analysis improvements in SK.\
The possible hardware improvements include an increase of the horn current from 
$\pm$250 kA to $\pm$320 kA. This will lead up to a 10\% more neutrino flux at the far detector.\
The analysis improvements include the expansion of the SK fiducial volume and/or adding new SK event selections.
As an example, adding CC1$\pi$ events will increase the event sample at the far detector by 14$\%$ 

According to the projections showed in Fig.\ref{fig:IntPOT} and in Fig.\ref{fig:MR-Power}, T2K will be able to collect neutrino events corresponding to
$20\times 10^{21}$ p.o.t. ($25\times 10^{21}$ p.o.t) before the year 2026 (2028) (red dots) or 2024 (2025) (blue dots).\\

Tables \ref{tab:78-25_nue},\ref{tab:78-25_numu} \cite{34,35} show the expected numbers of  $\nu_e$ or $\bar{\nu}_e$ appearance events and $\nu_\mu$ or $\bar{\nu}_\mu$ disappearance events 
for two different values of $\delta_{CP}$ (-90\(\degree\),0\(\degree\))
at $7.8\times10^{21}$~p.o.t. and $25\times10^{21}$~p.o.t. respectively.\
A combination of 50$\%$ $\nu$ + 50$\%$ $\bar{\nu}$ mode beam running is assumed.\\

\begin{table}[htbp] 
\caption{Expected numbers of  $\nu_e$ or $\bar{\nu}_e$ appearance events for two different values of $\delta_{CP}$ (-90\(\degree\),0\(\degree\))
at $7.8\times10^{21}$~p.o.t. and $25\times10^{21}$~p.o.t. respectively.\
A combination of 50$\%$ $\nu$ + 50$\%$ $\bar{\nu}$ mode beam running is assumed.}
\begin{center}

\begin{tabular}{  c | c | c | c | c | c  } \hline 
 &  &  $\nu_e$ Signal & $\nu_e$ bkg. &  $\bar{\nu}_e$ Signal & $\bar{\nu}_e$ bkg. \\ \hline
7.8E21 POT & 0\(\degree\) & 98.2 & 26.8 & 25.6 & 16.3  \\ 
 & -90\(\degree\) & 121.4 & 26.4 & 19.0 & 17.2  \\ \hline 
25E21 POT & 0\(\degree\) & 314 & 85.9 &  82.1 &  52.2  \\ 
&  -90\(\degree\) & 389 &  84.6 & 60.9 &  55.1  \\ \hline 
\end{tabular}
\end{center} 
\label{tab:78-25_nue} \end{table}

\begin{table}[htbp] 
\caption{Expected numbers of  $\nu_\mu$ or $\bar{\nu}_\mu$ disappearance events 
assuming or not the oscillation hypothesis 
at $7.8\times10^{21}$~p.o.t. and $25\times10^{21}$~p.o.t. respectively.\
A combination of 50$\%$ $\nu$ + 50$\%$ $\bar{\nu}$ mode beam running is assumed.}
\begin{center}

\begin{tabular}{  c | c | c | c } \hline 
 &  &   $\nu_{\mu}$  - model & $\bar{\nu}_\mu$ - mode \\ \hline
7.8E21 POT & w/o oscillation  & 2648 & 1007  \\ 
 & w/ oscillation & 741 & 342 \\ \hline 
25E21 POT & w/o oscillation & 8519  & 3228 \\ 
&  w/ oscillation & 2375 &  1096  \\ \hline 
\end{tabular}
\end{center} 
\label{tab:78-25_numu} \end{table}

\subsection{T2K Sensitivities to the oscillation parameters with $20 \times 10^{21}$ p.o.t.}

At the beginning of January 2016 the T2K collaboration has submitted an EoI (Expression of Interest) \cite{42} to the Japanese PAC Commitee, aiming to extend the T2K run to $20 \times 10^{21}$ p.o.t (T2K-II). 
For this study  T2K systematic errors are encoded into a covariance matrix with bins in reconstructed neutrino energy.\ Errors on the shape of the reconstructed energy spectra are taken into account.\
Two hypothesis have been considered on both reconstructed ${\nu}_e$ appearance and ${\nu}_\mu$ disappearance events: the current (2016) T2K systematic error on the far detector prediction
\footnote{the total systematic error on the far detector prediction is now 6.8$\%$ \cite{32}} and a possible reduction to 4$\%$  that seems to be reachable in the next years by T2K. \\

\begin{figure}[h!]
\begin{center}
\psfig{figure=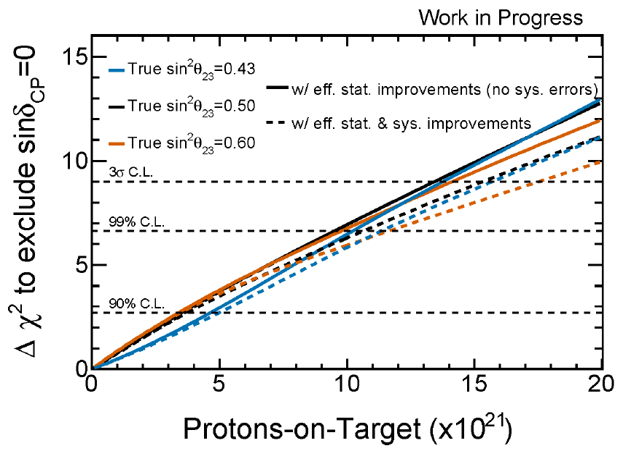,width=8cm}
\caption{Predicted $\Delta\chi^2$ for rejecting the
$\sin\delta_{CP}=0$ hypothesis versus p.o.t. assuming $\delta_{CP}$ = -90\(\degree\) and NH
for various true values of \(\sin^2\theta_{23}\) with or without  systematic errors.}
\label{fig:sens_251021}
\end{center}
\end{figure}

\begin{figure}[h!]
\begin{center}
\psfig{figure=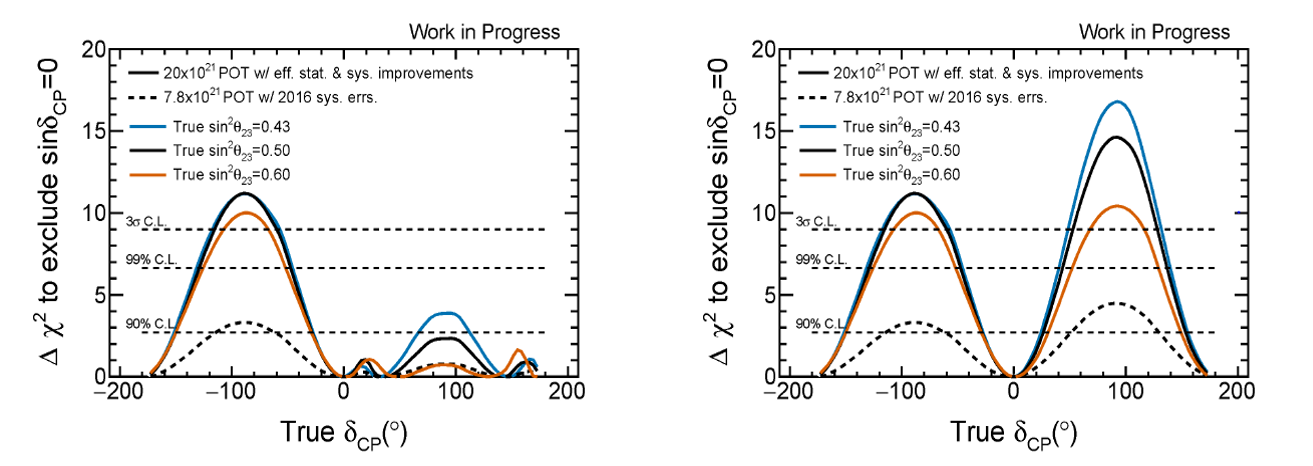,width=16cm}
\caption{ Sensitivities to CP violation as a function of true value of  \(\delta_{CP}\) for T2K  and extended T2K 
for various true values of \(\sin^2\theta_{23}\) assuming (right) or not (left) that the true MH is the normal MH.}
\label{fig:delta_251021}
\end{center}
\end{figure}

The T2K sensitivity to a non-zero $\sin(\delta_{CP})$ also depends on the true values of the oscillation parameters. 
In this case  all plots assume  \(\sin^22\theta_{13}\) = 0.085, \(\Delta m^2_{32}\) = $2.5 \times 10^{-3}$.\\
The updated horn current of $\pm$320 kA and a combination of 50$\%$ $\nu$ + 50$\%$ $\bar{\nu}$ mode beam running are also assumed.\\
In Fig.\ref{fig:sens_251021} the predicted $\Delta\chi^2$ for rejecting the
$\sin\delta_{CP}=0$ hypothesis is plotted versus p.o.t. (assuming $\delta_{CP}$ = -90\(\degree\)and true normal mass hierarchy (NH)) for various true values of \(\sin^2\theta_{23}\)  with or without  systematic errors.\
It is clear that  not only the improvement in statistics but also the reduction of the T2K systematic errors would be more than beneficial in the case of an extension of T2K.\\
Also the knowledge of the mass hierarchy is an important element, as it can be seen by looking at Fig.\ref{fig:delta_251021} where $\Delta\chi^2$ for rejecting the $\sin\delta_{CP}=0$ is plotted versus true $\delta_{CP}$. 
In fact, comparing the figure on the right (that assumes MH is known) with the figure on the left 
(that  assume MH is unknown), it is clear that the sensitivity to $\sin\delta_{CP}=0$  will be enhanced in the first case.
However several experiments (JUNO, NO$\nu$A, ORCA, PINGU) are expected or plan to determine
the mass hierarchy before or during the proposed period of T2K-II \cite{43,21,44,45}.  In this context the sensitivity shown in Fig. \ref{fig:delta_251021} (right) seems to illustrate a realistic scenario.\\

\section{Conclusions}
\label{Conclusion}

The T2K experiment, proposed in 2003 and 
approved in 2006 to collect data corresponding to $7.8\times 10^{21}$ protons-on-target (p.o.t.)
from a 30 GeV proton beam at the JPARC accelerator facility in Japan, 
is one of the most important players in the field of neutrino oscillations.\\
Build to search for \numu{}\goesto{}\nue{} appearance and  to make precision measurements
of oscillation parameters in \numu{} disappearance, it realized its first goal with just 
8.4$\%$ of the total approved p.o.t. and at the same time provided the most 
stringent constraints on $\sin^2(\theta_{23})$ obtained until now.\\
The T2K collaboration is now looking to the determination of the unknown 
CP-violating phase $\delta_{CP}$ and to more precise measurements of $\theta_{23}$ to determine the octant.\\ 
A re-evaluation of the expected sensitivity to the oscillation parameters that takes
into account the observation of the electron neutrino appearance was done considering two different scenarios: the approved data taking and a data exposure 3 times larger.\\
In the latter case, assuming a combination of 50$\%$ $\nu$ + 50$\%$ $\bar{\nu}$ mode beam running, 
it might be possible to obtain a 3 $\sigma$ measurement excluding  $\sin(\delta_{CP})= 0$ (assuming $\delta_{CP}$ = -90\(\degree\) and NH)  around the year 2025, before the next generation of neutrino experiments will start their operation.\\

\vspace{1cm}

\end{document}